\documentclass{aa}
\usepackage{graphicx,subfigure,rotating}
\usepackage[light,all]{}

\usepackage{natbib,references}
\bibpunct{(}{)}{;}{a}{}{,} 

\begin{document}
   \title{Simultaneous Optical and X-ray Observations of Flares and Rotational Modulation 
   on the RS CVn Binary\\HR 1099 (V711 Tau) from the MUSICOS 1998 Campaign\thanks{Based on observations obtained during the MUSICOS 98 
   MUlti-SIte COntinuous Spectroscopic campaign from Observatoire de Haute-Provence, France, Kitt Peak National Observatory, USA, ESO La
   Silla, Chile, Mt. Stromlo Observatory, Australia, Xinglong National Observatory, China, Isaac Newton Telescope, Spain, Laborat\'{o}rio 
   Nacional de Astrof\'{\i}sica, Brazil, and South African Astronomical Observatory, South Africa. Contemporaneous observations from Catania, Italy and Fairborn 
   Observatories, USA, and on data obtained with the Rossi X-ray Timing Explorer.}}


\author{D. Garc\'{\i}a-Alvarez\inst{1}
          \and B.H. Foing\inst{2}
	  \and D. Montes\inst{3}
	  \and J. Oliveira\inst{2,4,5}
	  \and J.G. Doyle\inst{1}
	  \and S. Messina\inst{6}
	  \and A. F. Lanza\inst{6}	  
	  \and M. Rodon\`o\inst{7}
	  \and J. Abbott\inst{8,9}
	  \and T. D. C. Ash\inst{10}
	  \and I. K. Baldry\inst{11,12}
	  \and T. R. Bedding\inst{11}
	  \and D.A.H. Buckley\inst{13}
	  \and J. Cami\inst{14}
	  \and H. Cao\inst{15}
	  \and C. Catala\inst{16}
	  \and K.P. Cheng\inst{17}
	  \and A. Domiciano de Souza Jr\inst{18}
	  \and J.-F. Donati\inst{16}
	  \and A.-M. Hubert\inst{19}
	  \and E. Janot-Pacheco\inst{20}
	  \and J. X. Hao\inst{15}
	  \and L. Kaper\inst{14}
	  \and A. Kaufer\inst{21}
	  \and N. V. Leister\inst{18}
	  \and J. E. Neff\inst{22}
	  \and C. Neiner\inst{14,19}
	  \and S. Orlando\inst{23}
	  \and S. J. O'Toole\inst{11}
	  \and D. Sch\"afer\inst{24,25}
	  \and S. J. Smartt\inst{24}
	  \and O. Stahl\inst{25}
	  \and J. Telting\inst{26}
          \and S. Tubbesing\inst{25}
}
          
\offprints{D. Garc\'{\i}a-Alvarez\\
\email{dga@star.arm.ac.uk}}

\institute{Armagh Observatory, College Hill, Armagh BT61 9DG N.Ireland
	\and Research Support Division, ESA RSSD, ESTEC/SCI-SR  postbus 299, 2200 AG Noordwijk, The Netherlands
	\and Dept de Astrof\'{\i}sica, Facultad de Ciencias F\'{\i}sicas, Universidad Complutense de Madrid, E-28040 Madrid, Spain
	\and Centro de Astrof\'{\i}sica da Universidade do Porto, Rua das Estrelas s/n, 4150-762 Porto, Portugal
	\and Astrophysics Group, School of Chemistry and Physics, Keele University, Staffordshire ST5 5BG, UK 
	\and Catania Astrophysical Observatory, Via S. Sofia, 78 I-95123 Catania, Italy
	\and Department of Physics and Astronomy, Catania University,  Via S. Sofia, 78  I-95123 Catania, Italy
	\and Isaac Newton Group of Telescopes, NWO, Apartado 321, 38700 Santa Cruz de La Palma, Spain
	\and Dept of Physics and Astronomy, University College London, Gower St., London WC1E 6BT, UK 
	\and Kildrummy Technologies, Mill Lane, Lerwick, Shetland, ZE1 0LX
	\and School of Physics, University of Sydney, NSW 2006, Australia
        \and Dept of Physics and Astronomy, Johns Hopkins University, Baltimore, USA
	\and South African Astronomical Observatory, PO Box 9, Observatory 7935, Cape Town, South Africa
	\and Astronomical Institute Anton Pannekoek, Univ of Amsterdam, 1098 SJ Amsterdam, The Netherlands
	\and Beijing Astronomical Observatory, Datun Road A20, Chaoyang District, Beijing 100012, P.R. China
	\and Laboratoire d'Astrophysique, Observatoire Midi-Pyrenees, 14 avenue Edouard Belin, F-31400 Toulouse, France
	\and California State University, Fullerton, CA, USA
	\and Instituto de Astronomia e Geofisica, Universidade de Sao Paulo, 1226/05508-900 Sao Paulo, Brazil
	\and GEPI/FRE K 2459 du CNRS, Observatoire de Paris-Meudon, France
	\and Observatoire de la C\^ote d'Azur, Dept FRESNEL, CNRS UMR 6528, 06130, Grasse, France
	\and European Southern Observatory, Alonso de Cordova 3107, Santiago 19, Chile
	\and Dept of Physics and Astronomy, College of Charleston, Charleston, SC 29424, USA
	\and Osservatorio Astronomico di Palermo "G.S. Vaiana", Piazza del Parlamento 1, I-90134 Palermo, Italy
	\and Institute of Astronomy, University of Cambridge, Madingley Road, Cambridge CB3 0HA, England
	\and Landessternwarte Heidelberg, K\"onigstuhl 12, 69117 Heidelberg, Germany
	\and Nordic Optical Telescope, Apartado 474, 38700 Santa Cruz de La Palma, Spain
}
   \date{Received, 2002 / accepted, 2002}

   \abstract{
We present simultaneous and continuous observations of the H$\alpha$, H$\beta$, 
He\,{\sc i} D$_{3}$, Na\,{\sc i} D$_{1}$,D$_{2}$ doublet and the Ca\,{\sc ii} H \& K lines for 
the RS CVn system 
HR 1099. The spectroscopic 
observations were obtained during the MUSICOS 1998 campaign involving several observatories 
and instruments,
both echelle and long-slit spectrographs. During this campaign, HR 1099 was observed almost
continuously for more than 8 orbits of $2\fd8$. Two large optical flares were observed, both showing an 
increase in the emission of H$\alpha$, Ca\,{\sc ii} H \& K, H$\beta$ and 
He\,{\sc i} D$_{3}$  and a strong filling-in of the Na\,{\sc i} D$_{1}$,D$_{2}$ doublet.
{ Contemporary photometric observations were carried out with 
the robotic telescopes APT-80 of Catania and Phoenix-25 of 
Fairborn Observatories. Maps of the distribution of the spotted regions 
on the photosphere of the binary components were derived 
using the Maximum Entropy and Tikhonov photometric regularization criteria}. Rotational modulation 
was observed in H$\alpha$ and He\,{\sc i} D$_{3}$ 
 in anti-correlation with the photometric light curves. Both
flares occurred at the same binary phase (0.85), suggesting that these events took place in the same 
active region. 
Simultaneous X-ray observations, performed by ASM on board RXTE, show several
flare-like events, some of which correlate well with the observed optical 
flares. Rotational modulation in the X-ray light 
curve has been detected with minimum flux when the less active G5\,V star was in 
front. A possible periodicity in the X-ray flare-like events was also found.
}

\maketitle

\markboth{D. Garc\'{\i}a-Alvarez et al.: Multi-wavelength Observations of HR 1099 (V711 Tau)}
{D. Garc\'{\i}a-Alvarez et al.: Multi-wavelength Observations of HR 1099 (V711 Tau)}

\section{Introduction}
RS CVn binary systems consist of a chromospherically active evolved 
star tidally locked to a main-sequence or sub-giant companion. Short  orbital periods of 
a few days are typically observed. The RS CVn high level of activity has been 
measured across the {entire spectrum} and, for fast rotators, it 
approaches the saturation limits for chromospheric, transition region and 
coronal emission. One of the striking aspects of these systems is their propensity to
flare \citep{Doyle92}. Moreover,  RS CVns show optical photometric variations which are 
believed to arise from the rotational modulation of photospheric spots 
\citep{Rodono86}, large scale versions of dark solar spots, that provide 
evidence for large scale magnetic fields. The typical 
energies {of the atmospheric magnetic fields} in these systems are up to several orders of 
magnitude larger 
than on the Sun, thus allowing us to observe a range of energetic phenomena 
not occurring on the Sun. Short-period RS CVn-like systems, through their 
rotational modulation, can then provide information on the morphology and 
three-dimensional spatial distribution of spots in stellar atmospheres.

The system observed in this campaign is HR 1099 (V711 Tau, 
$03^h36^m47^s$ $+00^o35'16''$  (J2000), { V=5.64, B$-$V=0.92}) a close 
double-lined spectroscopic binary. At a distance of 29 pc (The {\it{Hipparcos}} and Tycho 
Catalogues \citep{ESA97}), 
HR 1099 is the nearest 
and brightest of the classical RS CVns with a  
K1\,IV primary and a G5\,V secondary tidally locked in a 
$2\fd8$ orbit.  \citet{Fekel83} provided the orbital 
parameters of the binary and obtained  masses, radii, 
and spectral types for both components. The K sub-giant nearly fills its Roche lobe 
and is by far the most ``active'' of the two {components} \citep{AyresLinsky82,Robinson96}.
{Evolutionary models 
suggest that mass transfer from the K primary onto the G secondary may begin 
within $10^{7}$ years \citep{Fekel83}.} 
The primary exhibits conspicuous signatures of chromospheric activity, such as strong and variable 
Ca\,{\sc ii} H \& K and 
H$\alpha$ emission \citep{Rodono87,Dempsey96,Robinson96}, which is due to its tidally 
induced 
rapid rotation, combined with the deepened convection zone of a post-main-
sequence envelope. \citet{Montes97} revealed 
broad and variable wings of the H$\alpha$ chromospheric line on the primary star.

\begin{figure}[htbp!]
   \centering
   \resizebox{\hsize}{!}{\includegraphics{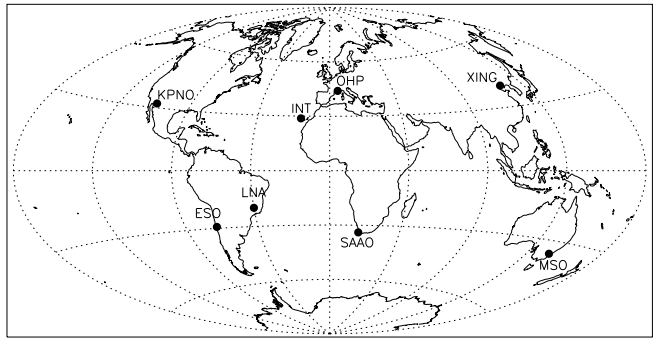}}
   \caption{The sites involved in the MUSICOS 98 campaign: INT, OHP, XING 
   (Xinglong), KPNO (Kitt Peak), MSO (Mt. Stromlo), LNA, ESO and SAAO.}
   \label{FCoverage}
\end{figure}

The G5 dwarf is probably also active, having a sufficiently deep convection zone and 
fast rotation to host an efficient dynamo. In particular, due to the presence of spots, G-type 
stars with rotation periods of about 3 days {are expected to} show V-band light curve 
amplitudes up to 0.10 mag, as can 
be inferred from the rotation-activity relations at photospheric levels \citep{Messina01}. However, 
because of the small luminosity ratio {in the V-band} (L$_{\rm G5V}$/L$_{\rm K1IV}\simeq 
0.27$), the contribution 
from the G5 component to the observed magnetic activity manifestations {in the optical band
 is significantly smaller than from the K1\,IV primary.}

HR 1099 is one of the few RS CVn systems, along with UX Ari, II Peg, and DM 
UMa, that shows H$\alpha$ {constantly} in emission. Optical light curves have shown, via 
Doppler imagery, the presence of a 
long-lived {($>$11 yr)} polar spot, together with transient ($<$1 yr) low-latitude 
spots on the surface of the active K star \citep{Vogt99}. Strong magnetic fields
of the order of 1000 G have also been detected 
on HR 1099 using the Zeeman Doppler Imaging technique 
{\citep{Donati90,Donati92,Donati99}.}

\citet{White78} reported the first X-ray outburst in coincidence with a large radio 
flare on HR 1099. \citet{vandenOord88} argued that {the low amplitude 
variability of the HR 1099 light curve on time-scales of tens of minutes would indicate
 flare-like heating}. This is in-keeping with the earlier work of \citet{Doyle85} 
 who suggested that the
 corona of all late-type stars were heated via a flare-like process. 
 \citet{Pasquini89} suggested coronal 
temperatures of 3 MK and 25 MK based on spectral fits to the {X-ray} data. Several 
authors \citep{Agrawal88,Drake94,Audard01} have reported periodic changes in the
{X-ray and UV} light curves, which they attributed to rotational modulation of a 
starspot-associated bright coronal region. \citet{Dempsey96} recorded, in coordinated IUE and 
GHRS observations, several flare enhancements in UV lines, one of which lasted 
more than a day. \citet{Wood96} suggested a relationship between the broad wings in 
UV emission lines of HR 1099 and micro-flaring in its chromosphere and 
transition region. \citet{ Brown97} reported variability in high-energy bands  with ASCA, 
RXTE, and EUVE: numerous small flares and a giant event lasting 3 days. 

\begin{table*}[htbp!]
\begin{center}
\caption{The sites and instruments involved {in the spectroscopic} observation of HR 1099 
during the MUSICOS 98 campaign and some of their most important characteristics: number of nights allocated
at each site and instrument, resolving power, number of spectral orders, wavelength coverage (nm) and number of spectra obtained.}
\begin{tabular}{lccccccc}
\hline
\hline\\
Site & Telescope & Spectrograph & Number & Resolving & Number & Wavelength & Number\\
     &           &              & Nights & Power     & Orders & coverage (nm) & Spectra\\
\hline\\ 
1. OHP, France   & 1.5 m & Aurelie & 6 & 22\,000 & 1 & 652-672 & 37\\
2. OHP, France & 1.9 m & Elodie & 8 & 43\,000 & 67 & 390-690  & 7\\
3. KPNO, USA & 0.9 m & Echelle & 10 & 65\,000 & 23 & 530-700 & 20\\
4. ESO, Chile & 0.9 m & HEROS & 11 & 20\,000 & 62 & 350-560 & 34\\
           &       &       &    &         & 33 & 580-865 & 34\\
5. Mt.Stromlo, Australia & 1.9 m & Echelle & 5 & 35\,000 & 43 & 480-680 & 17\\
6. Xinglong, China & 2.2 m & Echelle & 10 & 35\,000  & 35 & 550-850 & 10\\
7. INT, Spain & 2.5 m & ESA-MUSICOS & 9 & 35\,000 & 58 & 400-680 & 13 \\
8. LNA, Brazil & 1.6 m & Coud\'{e} & 6 & 60\,000 & 1 & 663-672 & 5\\   
9. SAAO, South Africa & 1.9 m & Giraffe & 3 & 36\,500 & 52 & 430-700 & 3 \\
\hline \\
\end{tabular}
\end{center}
\end{table*}

\begin{figure*}[htbp!]
   \centering
\begin{turn}{90}
 \includegraphics[width=14cm]{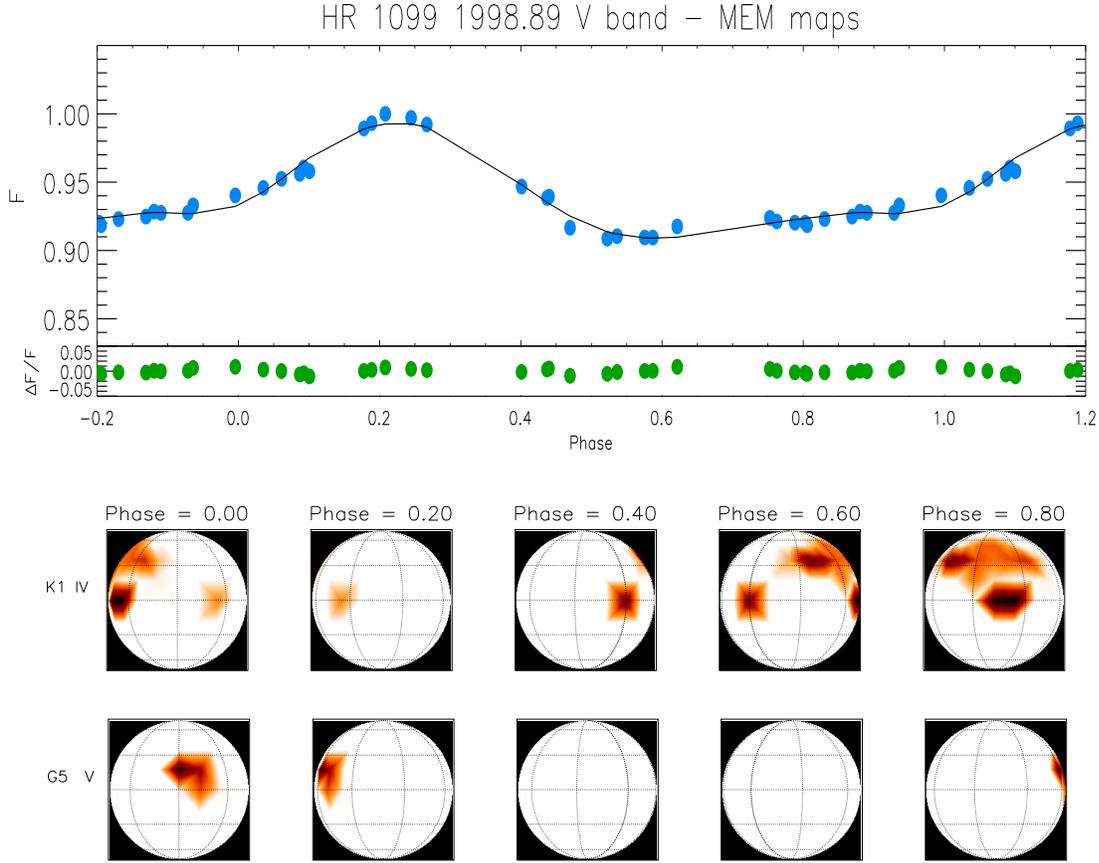}
\end{turn}
\vspace*{-1.0cm}
   \caption{Top panel: V-band light curve (filled dots) of HR 1099 in late November/early December
   1998 fitted by the 
   Maximum Entropy spot model (continuous line). The flux was normalized to the brightest 
magnitude (V$_{\rm unsp}=5.744$ at phase=0.21) and phases were computed according to 
Eq.~(\ref{ephemeris}). 
The residuals ($\Delta F / F$) between the observed and synthetized light curves are also plotted 
vs. phase. Bottom panel: Maps of  the 
distribution along stellar longitude of the spot filling factors at five rotation phases. Spots located at 
latitude below $\simeq -33^{\circ}$  cannot contribute to the flux because  the  inclination of the 
star's rotation axis is $ 
33^{\circ}$.}
   \label{FigPhotometry1}
\end{figure*}

\begin{figure*}[htbp!]
   \centering
\begin{turn}{90}
 \includegraphics[width=14cm]{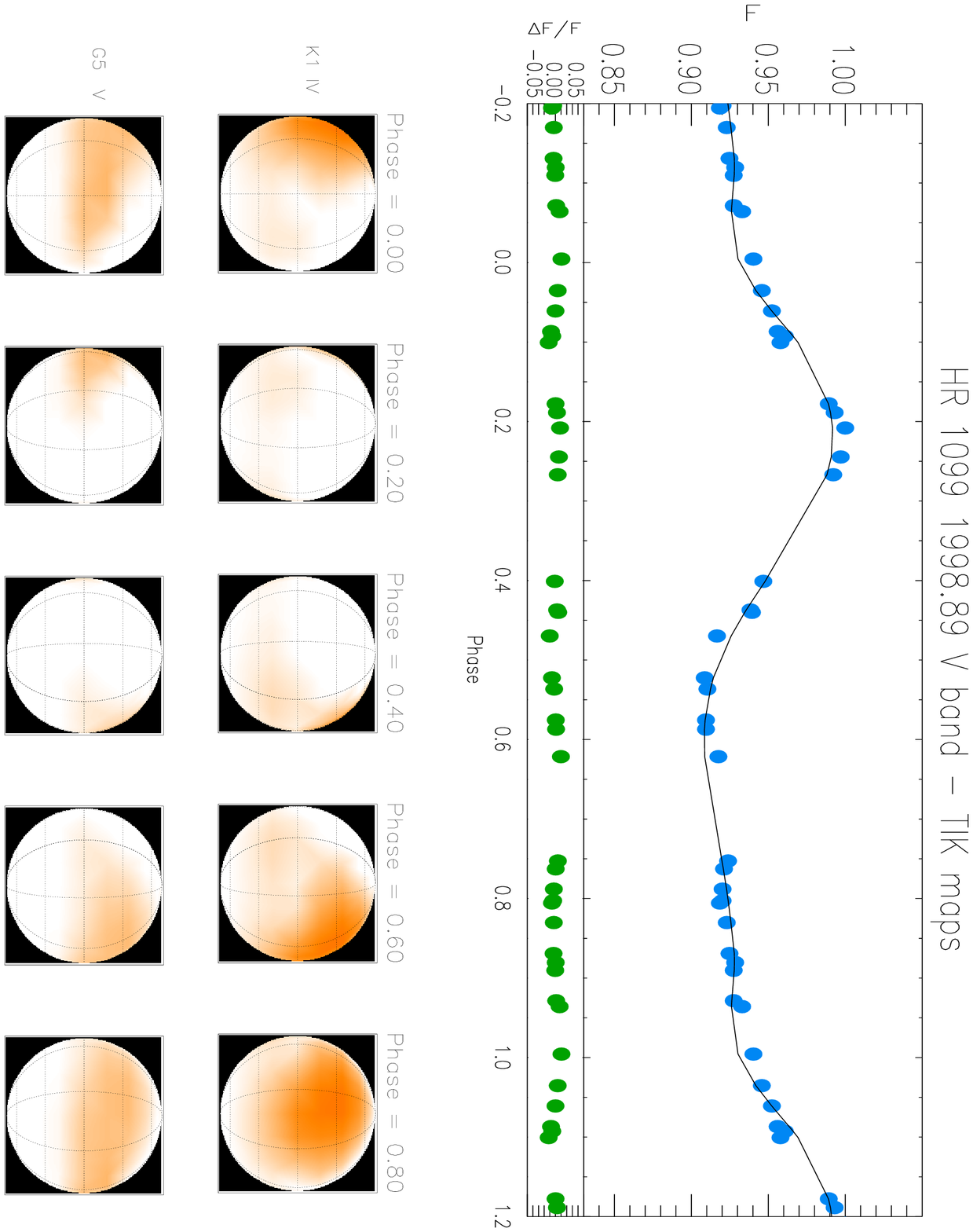}
\end{turn}
\vspace*{-1.0cm}
   \caption{The same as in Fig.~\ref{FigPhotometry1}  fitted by  the Tikhonov spot model and the 
Tikhonov 
photospheric maps. The presence of spots in the inaccessible regions is an artifact of the Tikhonov 
regularization that correlates the filling factors of nearby pixels.}
   \label{FigPhotometry2}
\end{figure*}

Many scientific programs, most of them linked to stellar physics (such as flare 
monitoring, stellar rotational modulation, surface structures, Doppler imaging) 
requires continuous spectroscopic coverage over several days. 
MUSICOS\footnote{http://www.ucm.es/info/Astrof/MUSICOS.html} (MUlti-SIte COntinuous 
Spectroscopy) is an international project for setting up 
a network of high resolution spectrographs coupled to telescopes well distributed 
around the world. During the {MUSICOS 1989} campaign on HR 1099, \citet{Foing94} 
obtained complete phase coverage for Doppler imaging of spots. 
They observed the modulation of the Ca\,{\sc ii} K line profile due to 
chromospheric plage-like regions. They also observed two white-light flares (one of 
them was the largest optical flare reported on an RS CVn system) with typical 
rise and decay times of 60-90 min, and with a remarkable spectral 
dynamic signature in H$\alpha$. The fifth MUSICOS campaign (MUSICOS 98), the largest campaign organised to date, 
involved {13 telescopes} to address 6 distinct science programs. Fig.~\ref{FCoverage} shows 
the distribution of the 
participating sites. The main goals of the campaign for HR 1099 were to monitor for 
flare events, to probe the chromospheric line variability in order to diagnose the energetics 
and dynamics in active regions, and to produce photospheric Doppler images.

In this paper we present the 
results of the MUSICOS 98 campaign on HR 1099. Observations and data analysis 
are described in Sect.~2; in Sect.~3 we present the results, and in Sect.~4 we 
analyse the flares on HR 1099. The conclusions are given in Sect.~5. 
Note that the fragmentary results from a preliminary analysis published in conference proceedings 
\citep{Garcia-Alvarez01a,Garcia-Alvarez01b} are superseded by the final ones presented herein. 

\begin{table*}[]
\begin{center}
\caption{Log of the HR 1099 {spectroscopic} observations {during} the MUSICOS 98 campaign.}
\scriptsize{
\begin{tabular}{lccrclccrc}
\hline
\hline\\
Date & JD         & Phase   & Exp.  & Site     &    Date & JD         & Phase   & Exp.  & Site     \\
     & 2451100.0+ &  & (sec.)    &      &        & 2451100.0+ & mid exp & (sec.)
 &              \\
\hline\\
1998 Nov 21    & 39.390  &  0.6994  &    600   & 1 &  1998 Nov 30    & 48.046  &  0.7498  
&        800  & 5 \\
1998 Nov 21    & 39.430  &  0.7136  &    600   & 1 &  1998 Nov 30    & 48.147  &  0.7853  
&         800 & 5 \\
1998 Nov 21    & 39.494  &  0.7364  &    600   & 1 &  1998 Nov 30    & 48.333  &  0.8509  
&        1200 & 1 \\
1998 Nov 22    & 40.328  &  0.0300  &     1200 & 1 &  1998 Nov 30    & 48.356  &  0.8592  
&        1200 & 1 \\
1998 Nov 22    & 40.363  &  0.0425  &     1200 & 1 &  1998 Dec 01    & 48.519  &  0.9167  
&        1800 & 4 \\
1998 Nov 22    & 40.475  &  0.0819  &     1200 & 1 &  1998 Dec 01    & 48.597  &  0.9441  
&        1800 & 4 \\
1998 Nov 23    & 40.574  &  0.1169  &     1200 & 1 &  1998 Dec 01    & 48.932  &  0.0621  
&        900  & 5 \\
1998 Nov 23    & 40.659  &  0.1468  &     1200 & 1 &  1998 Dec 01    & 48.963  &  0.0731  
&        800  & 5 \\
1998 Nov 23    & 41.303  &  0.3739  &    600   & 2 &  1998 Dec 01    & 49.097  &  0.1203  
&        800  & 5 \\
1998 Nov 23    & 41.329  &  0.3829  &     1200 & 1 &  1998 Dec 01    & 49.144  &  0.1367  
&         800 & 5 \\
1998 Nov 23    & 41.367  &  0.3961  &     1200 & 1 &  1998 Dec 01    & 49.438  &  0.2402  
&    600   & 1 \\
1998 Nov 23    & 41.423  &  0.4160  &    600   & 2 &  1998 Dec 02    & 49.519  &  0.2691  
&        1800 & 4 \\
1998 Nov 23    & 41.470  &  0.4326  &     1200 & 1 &  1998 Dec 02    & 49.612  &  0.3017  
&        1800 & 4 \\
1998 Nov 24    & 41.569  &  0.4676  &     1200 & 1 &  1998 Dec 02    & 49.934  &  0.4152  
&        900  & 5 \\
1998 Nov 24    & 41.585  &  0.4732  &     1200 & 3 &  1998 Dec 02    & 50.011  &  0.4424  
&        1200 & 6 \\
1998 Nov 24    & 41.654  &  0.4975  &     1200 & 1 &  1998 Dec 02    & 50.074  &  0.4647  
&        1200 & 6 \\
1998 Nov 24    & 41.711  &  0.5175  &    900   & 3 &  1998 Dec 02    & 50.185  &  0.5036  
&        1200 & 6 \\
1998 Nov 24    & 41.853  &  0.5674  &     1200 & 3 &  1998 Dec 03    & 50.516  &  0.6203  
&        1800 & 4 \\
1998 Nov 24    & 41.959  &  0.6049  &      1200& 3 &  1998 Dec 03    & 50.608  &  0.6528  
&        1800 & 4 \\
1998 Nov 24    & 42.319  &  0.7316  &    600   & 2 &  1998 Dec 03    & 50.975  &  0.7821  
&    900   & 5 \\
1998 Nov 24    & 42.327  &  0.7346  &     1200 & 1 &  1998 Dec 03    & 51.007  &  0.7933  
&        800  & 5 \\
1998 Nov 24    & 42.351  &  0.7429  &     1200 & 1 &  1998 Dec 03    & 51.066  &  0.8141  
&        800  & 5 \\
1998 Nov 24    & 42.451  &  0.7781  &     1200 & 1 &  1998 Dec 03    & 51.126  &  0.8354  
&        800  & 5 \\
1998 Nov 25    & 42.543  &  0.8107  &     1200 & 1 &  1998 Dec 03    & 51.184  &  0.8557  
&        800  & 5 \\
1998 Nov 25    & 42.584  &  0.8251  &     1200 & 3 &  1998 Dec 03    & 51.287  &  0.8922  
&  1800    & 9 \\
1998 Nov 25    & 42.627  &  0.8403  &     1200 & 1 &  1998 Dec 03    & 51.346  &  0.9127  
&  1800    & 9 \\
1998 Nov 25    & 42.735  &  0.8782  &     1200 & 3 &  1998 Dec 03    & 51.446  &  0.9480  
& 900      & 9 \\
1998 Nov 25    & 42.838  &  0.9145  &     1200 & 3 &  1998 Dec 04    & 51.514  &  0.9720  
&        1800 & 4 \\
1998 Nov 25    & 42.938  &  0.9499  &      1200& 3 &  1998 Dec 04    & 51.610  &  0.0057  
&        1800 & 4 \\
1998 Nov 26    & 43.510  &  0.1516  &     1200 & 1 &  1998 Dec 04    & 52.035  &  0.1557  
&        1200 & 6 \\
1998 Nov 26    & 43.594  &  0.1812  &     1800 & 4 &  1998 Dec 04    & 52.062  &  0.1653  
&        1200 & 6 \\
1998 Nov 26    & 43.622  &  0.1907  &     1200 & 1 &  1998 Dec 04    & 52.168  &  0.2025  
&         1200& 6 \\
1998 Nov 26    & 43.677  &  0.2103  &     1800 & 4 &  1998 Dec 04    & 52.247  &  0.2304  
&        1200 & 6 \\
1998 Nov 26    & 43.759  &  0.2392  &     1200 & 3 &  1998 Dec 05    & 52.513  &  0.3241  
&        1800 & 4 \\
1998 Nov 26    & 43.847  &  0.2700  &     1200 & 3 &  1998 Dec 05    & 52.598  &  0.3540  
&        1200 & 7 \\
1998 Nov 26    & 43.967  &  0.3124  &     1200 & 3 &  1998 Dec 05    & 52.606  &  0.3566  
&        1800 & 4 \\
1998 Nov 26    & 43.982  &  0.3177  &     1200 & 3 &  1998 Dec 05    & 53.088  &  0.5265  
&        1200 & 6 \\
1998 Nov 26    & 43.997  &  0.3231  &     1200 & 3 &  1998 Dec 06    & 53.514  &  0.6767  
&        1800 & 4 \\
1998 Nov 26    & 44.306  &  0.4318  &     1200 & 2 &  1998 Dec 06    & 53.614  &  0.7120  
&        1800 & 4 \\
1998 Nov 26    & 44.324  &  0.4384  &     1200 & 1 &  1998 Dec 06    & 54.479  &  0.0169  
&        1200 & 7  \\
1998 Nov 26    & 44.345  &  0.4457  &     1200 & 1 &  1998 Dec 07    & 54.511  &  0.0282  
&        1800 & 4  \\
1998 Nov 26    & 44.454  &  0.4842  &     1200 & 1 &  1998 Dec 07    & 54.606  &  0.0617  
&        1800 & 4  \\
1998 Nov 27    & 44.513  &  0.5050  &     1800 & 4 &  1998 Dec 07    & 54.669  &  0.0837  
&   1500   & 7  \\
1998 Nov 27    & 44.581  &  0.5289  &     1800 & 4 &  1998 Dec 07    & 55.108  &  0.2384  
&        1200 & 6  \\
1998 Nov 27    & 44.629  &  0.5458  &     1200 & 3 &  1998 Dec 07    & 55.170  &  0.2604  
&         1200& 6  \\
1998 Nov 27    & 44.644  &  0.5510  &     1200 & 3 &  1998 Dec 07    & 55.347  &  0.3228  
&   900    & 7  \\
1998 Nov 27    & 44.659  &  0.5563  &     1200 & 3 &  1998 Dec 07    & 55.444  &  0.3571  
&    600   & 8  \\
1998 Nov 27    & 44.783  &  0.5999  &     1200 & 3 &  1998 Dec 07    & 55.469  &  0.3659  
&    600   & 8  \\
1998 Nov 27    & 44.798  &  0.6053  &     1200 & 3 &  1998 Dec 07    & 55.491  &  0.3735  
&        1200 & 7  \\
1998 Nov 27    & 44.812  &  0.6104  &     1200 & 3 &  1998 Dec 08    & 55.514  &  0.3815  
&        1800 & 4  \\
1998 Nov 27    & 44.941  &  0.6557  &      1200& 3 &  1998 Dec 08    & 55.515  &  0.3820  
&    600   & 8  \\
1998 Nov 27    & 45.328  &  0.7923  &     1200 & 2 &  1998 Dec 08    & 55.612  &  0.4160  
&        1800 & 4  \\
1998 Nov 27    & 45.345  &  0.7981  &     1200 & 1 &  1998 Dec 08    & 56.330  &  0.6691  
&    900   & 7  \\
1998 Nov 27    & 45.462  &  0.8392  &     1200 & 1 &  1998 Dec 08    & 56.487  &  0.7246  
&        1200 & 7  \\
1998 Nov 28    & 45.513  &  0.8571  &     1800 & 4 &  1998 Dec 09    & 56.517  &  0.7349  
&        1800 & 4  \\
1998 Nov 28    & 45.549  &  0.8698  &     1200 & 1 &  1998 Dec 09    & 56.543  &  0.7442  
&    600   & 8  \\
1998 Nov 28    & 45.632  &  0.8992  &     1200 & 1 &  1998 Dec 09    & 56.550  &  0.7466  
&    600   & 8 \\
1998 Nov 28    & 46.324  &  0.1432  &   1500   & 2 &  1998 Dec 09    & 56.558  &  0.7493  
&    600   & 8 \\
1998 Nov 28    & 46.337  &  0.1476  &     1200 & 1 &  1998 Dec 09    & 56.613  &  0.7689  
&        1800 & 4 \\
1998 Nov 28    & 46.359  &  0.1554  &     1200 & 1 &  1998 Dec 09    & 57.364  &  0.0335  
&    900   & 7 \\
1998 Nov 28    & 46.454  &  0.1889  &     1200 & 1 &  1998 Dec 10    & 57.517  &  0.0873  
&         1800& 4 \\
1998 Nov 29    & 46.523  &  0.2132  &     1800 & 4 &  1998 Dec 10    & 57.524  &  0.0897  
&        900  & 7 \\
1998 Nov 29    & 47.017  &  0.3872  &    900   & 5 &  1998 Dec 10    & 57.610  &  0.1203  
&         1800& 4 \\
1998 Nov 29    & 47.031  &  0.3921  &    900   & 5 &  1998 Dec 10    & 58.302  &  0.3641  
&        900  & 7 \\
1998 Nov 29    & 47.156  &  0.4361  &     900  & 5 &  1998 Dec 10    & 58.341  &  0.3778  
&        900  & 7 \\
1998 Nov 29    & 47.374  &  0.5129  &   1500   & 2 &  1998 Dec 10    & 58.374  &  0.3893  
&         1200& 7 \\
1998 Nov 29    & 47.397  &  0.5210  &     1200 & 1 &  1998 Dec 11    & 58.515  &  0.4389  
&        1800 & 4 \\
1998 Nov 30    & 47.513  &  0.5619  &     1800 & 4 &  1998 Dec 11    & 58.610  &  0.4727  
&        1800 & 4 \\
1998 Nov 30    & 47.542  &  0.5724  &     1200 & 1 &  1998 Dec 11    & 58.627  &  0.4786  
&    900   & 7 \\
1998 Nov 30    & 47.586  &  0.5878  &     1800 & 4 &  1998 Dec 12    & 59.511  &  0.7901  
&        1800 & 4 \\
1998 Nov 30    & 47.635  &  0.6052  &     1200 & 1 &  1998 Dec 12    & 59.612  &  0.8256  
&        1800 & 4 \\
1998 Nov 30    & 47.929  &  0.7087  &     900  & 5 &  1998 Dec 13    & 60.516  &  0.1442  
&        1800 & 4 \\
1998 Nov 30    & 47.967  &  0.7219  &    900   & 5 &  1998 Dec 13    & 60.644  &  0.1893  
&        1800 & 4 \\
\hline \\
\end{tabular}
}
\end{center}
\end{table*}


\begin{table}[htbp!]
\begin{center}
\caption{Log of RTXE observations.}
\begin{tabular}{lcc}
\hline
\hline\\
Date & No. exp.  & Exp. time  \\
     &   & (sec.)   \\
\hline\\
1998 Nov 22   & 9   &90	\\
1998 Nov 23   & 13  &90	\\
1998 Nov 24   &  3  &90	\\
1998 Nov 25   & 12  &90	\\
1998 Nov 26   &  6  &90	\\
1998 Nov 27   & 16  &90	\\
1998 Nov 28   & 11  &90	\\
1998 Nov 29   & 17  &90	\\
1998 Nov 30   & 13  &90	\\
1998 Dec 01   &  7  &90	\\
1998 Dec 02   & 24  &90	\\
1998 Dec 03   & 33  &90	\\
1998 Dec 04   & 11  &90	\\
1998 Dec 05   & 11  &90	\\
1998 Dec 06   &  4  &90	\\
1998 Dec 07   & 22  &90	\\
1998 Dec 08   & 13  &90	\\
1998 Dec 09   & 11  &90	\\
1998 Dec 10   & 21  &90	\\
1998 Dec 11   & 11  &90	\\
1998 Dec 12   & 12  &90	\\
\hline\\      	     
\end{tabular}
\end{center}
\end{table}

\section{Observations and Data}
\subsection{MUSICOS 98 campaign: Optical Spectroscopy}
The spectroscopic observations were obtained from 21 November
to 13 December 1998 during the MUSICOS 98 campaign. It involved eight northern and southern 
sites, namely: 
Observatoire de Haute-Provence (OHP), Kitt Peak National Observatory (KPNO), 
European Southern Observatory (ESO), Mt. Stromlo Observatory (MSO), Xinglong National 
Observatory, Isaac Newton 
Telescope (INT), Laborat\'{o}rio Nacional de Astrof\'{\i}sica (LNA) and South African 
Astronomical Observatory (SAAO), using both echelle and long-slit spectrographs.
A summary of the sites and instruments involved in the campaign and some of 
their most important characteristics are given in Table~1. 

The spectra were extracted using the standard reduction procedures in the 
NOAO packages IRAF.\footnote{IRAF is distributed by the National Optical 
Astronomy Observatories, which is operated by the Association of Universities 
for Research in Astronomy, Inc., under cooperative agreement with the National
Science Foundation.} For OHP, HEROS and Xinglong observations, 
the MIDAS\footnote{ESO-MIDAS is the acronym for the European 
Southern Observatory Munich Image Data Analysis System, which is developed and
maintained by the European Southern Observatory.} 
package was used. Background subtraction and flat-field correction using 
exposures of a tungsten lamp were applied. The wavelength calibration was 
obtained by taking spectra of a Th--Ar lamp. {A first-order spline cubic fit to 
some 45 lines achieved a nominal wavelength calibration accuracy which ranged 
from 0.06 to 0.11 \AA.  The spectra were normalized by a low-order 
polynomial fit to the observed continuum. Finally, for the spectra affected by water lines, a 
telluric correction was applied. 

Apart from the 
chromospheric activity indicators, the spectra also include many lines important 
for spectral classification (i.e. Ca\,{\sc i} triplet 6102, 6122, 6162 \AA\ and Na\,{\sc i} doublet 8183.3,8194.8 \AA) 
and temperatures calibration purposes 
(i.e. V\,{\sc i} 6251.8 \AA\ and Fe\,{\sc i} 6252.6 \AA), 
as well as other lines normally used for the application of  the Doppler imaging technique (i.e. Fe\,{\sc i} 6411.7 \AA, 
Fe\,{\sc i} 6430.9 \AA\ and Ca\,{\sc i} 6439.1 \AA). 
A summary of the observations, including Julian date, phase at 
mid-exposure and site (refer to Table~1), is given in Table~2.{ To produce a phase-folded light 
curve we adopted 
the ephemeris given in Eq.~(\ref{ephemeris})}.

{Appropriate data have been obtained to apply Doppler Imaging technique, based on the photospheric 
lines, to study the connection between spots, chromospheric emission and flares. 
These results will be presented in a separate paper \citep{Garcia-Alvarez03}.}

\subsection{RXTE X-ray Observations}
The X-ray observations were obtained with the all-sky monitor (ASM) detector 
on board {the} \textit{Rossi X-ray Timing Explorer} (RXTE) \citep{Levine96}. The ASM 
consists of three similar scanning shadow cameras, sensitive to {X-rays} in an 
energy band of approximately 2-12 \rm{keV}, which perform sets of 90-second 
pointed observations (``dwells'') so as to cover 80\% of the sky every 90 
minutes. The analysis presented here makes use of light curves obtained from individual 
dwell data. Light curves are available in four energy bands: 
A (1.3-3.0\, \rm{keV}), B (3.0-4.8\, \rm{keV}), C (4.8-12.2\, 
\rm{keV}) and S (1.3-12.2\, \rm{keV}). Around 15 individual ASM dwells of HR 
1099 were observed daily by RXTE, during MUSICOS 98. A summary of 
the observations, including date, number of dwell data and exposure time are  
given in Table~3. The data were binned in four-hour intervals, which is 
approximately 6\% of the orbital period of HR 1099.

\subsection{Photometry}
{ During the MUSICOS 98 campaign, ground-based photoelectric observations were obtained
with two different Automatic Photoelectric
Telescopes (APTs):  a) the 
0.80-m APT-80 of Catania Astrophysical Observatory (CAO)  on Mt.\,Etna, Italy 
\citep{Rodono01}; 
b) the  0.25-m T1 {\it Phoenix} APT of Fairborn Observatory at Washington 
Camp, AZ, USA, that is managed as a multiuser telescope \citep{Boyd84,Seeds95}. Both 
telescopes 
are equipped with standard Johnson UBV filters and uncooled photoelectric 
photometers.
HR\,1099 {has been} regularly monitored since 1988 by the  {\it Phoenix}  \citep{Rodono92} 
and since 1992 also by the Catania APT-80 \citep{Cutispoto95}. 
Although repeated observations were scheduled on each night during the campaign to obtain complete 
rotation phase coverage, bad weather conditions resulted in light curve gaps around phases 0.30 
and 0.70.


\begin{table*}[htbp!]
\begin{center}
\caption{Radial velocities for the primary and secondary components of HR 1099, 
using the Fe\,{\sc i} 6430.84 \AA\ and Ca\,{\sc i}  6439.07 \AA\ lines, as 
measured from ESO-HEROS spectra.}
\begin{tabular}{cccccc}
\hline
\hline\\
JD & Orbital  & \multicolumn{2}{c} {$v_{r}$(km\,s$^{-1}$) Fe\,{\sc i} 6430.8 \AA\ } & 
\multicolumn{2}{c} 
{$v_{r}$(km\,s$^{-1}$) Ca\,{\sc i} 6439.1 \AA\ } \\
\cline{3-4}
\cline{5-6}
2451100.0+ & Phase &  pri. &  sec. &  pri. &  sec. \\
\hline\\ 
43.5940 &0.1812  &     +34.16	  &  $-78.32$	&   +32.03   & $-77.83$ \\
43.6770 &0.2103  &     +33.50	  &  $-76.18$	&   +32.55   & $-77.47$ \\
44.5810 &0.5289  &    $-36.51$    &   +17.36	& $-36.03$  &	+15.79  \\
45.5130 &0.8571  &    $-47.00$    &   +15.04	& $-46.26$  &	+20.59  \\
46.5230 &0.2132  &     +33.18	  &  $-76.55$	&   +32.19   & $-77.47$ \\
47.5130 &0.5619  &    $-45.45$    &   +29.77	& $-46.39$  &	+26.64  \\
47.5860 &0.5878  &    $-52.74$    &   +34.48	& $-53.21$  &	+32.28  \\
49.5190 &0.2691  &     +28.56	  &  $-72.72$	&   +27.53   & $-72.90$ \\
49.6120 &0.3017  &     +25.99	  &  $-66.80$	&   +24.16   & $-65.76$ \\
50.5160 &0.6203  &    $-57.11$    &   +39.60	& $-59.26$  &	+40.37  \\
50.6080 &0.6528  &    $-62.80$    &   +44.75	& $-64.33$  &	+44.46  \\
52.5130 &0.3241  &     +22.42	  &  $-61.27$	&   +19.45   & $-61.26$ \\
52.6060 &0.3566  &     +17.32	  &  $-50.55$	&   +11.32   & $-51.53$ \\
53.5140 &0.6767  &    $-64.71$    &   +45.57	& $-66.10$  &	+44.85  \\
53.6140 &0.7120  &    $-65.64$    &   +44.59	& $-67.58$  &	+46.64  \\
54.6060 &0.0617  &     +19.05	  &  $-55.85$	&   +14.34   & $-56.33$ \\
55.5140 &0.3815  &     +08.37	  &  $-41.22$	&   +04.84   & $-41.94$ \\
56.5170 &0.7349  &    $-62.66$    &   +42.93	& $-64.65$  &	+44.64  \\
56.6130 &0.7689  &    $-61.77$    &   +40.40	& $-63.44$  &	+40.94  \\
57.5170 &0.0873  &     +23.24	  &  $-63.28$	&   +20.73   & $-64.19$ \\
57.6100 &0.1203  &     +27.11	  &  $-71.24$	&   +25.62   & $-71.17$ \\
59.5110 &0.7901  &    $-55.76$    &   +41.56	& $-58.51$  &	+38.05  \\
59.6120 &0.8256  &    $-52.60$    &   +31.27	& $-54.03$  &	+29.66  \\
60.5160 &0.1442  &     +32.30	  &  $-73.47$	&   +28.96   & $-74.52$ \\
60.6440 &0.1893  &     +32.98	  &  $-78.28$	&   +31.57   & $-78.86$ \\
\hline \\									   	 
\end{tabular}									   	 
\end{center}
\label{TRV}
\end{table*}

The Catania APT-80 observed HR\,1099 differentially with respect to the comparison star 
HD\,22796 (V=5.55, B$-$V=0.94) and the check star HD\,22819 (V=6.10, B$-$V=1.00). 
The {\it Phoenix} APT observed HR 1099 differentially with respect to the comparison star 
HD\,22484 (V=4.28, B$-$V=0.58), while HD\,22796 (V=5.56, B$-$V=0.94) served as a check  
star. After 
subtraction of sky background and correction for atmospheric extinction,
the instrumental differential magnitudes were converted into the standard Johnson UBV system. 
In order to obtain a homogeneous data set the measurements from the two telescopes were 
 first converted into differential values with respect to the same comparison star HD\,22796 and 
then 
converted into standard values.
Details on the observation techniques, reduction procedures and accuracy of the {\it Phoenix} 
and the Catania APT-80 photometry are given in \citet{Rodono92} and \citet{Cutispoto01}, 
respectively.

The V-band flux modulation shown by HR 1099 in 1998.89 (see top panel of 
Fig.~\ref{FigPhotometry1}) can be attributed to the presence of a 
stationary spot pattern, mainly on the photosphere of the K1\,IV primary, whose visibility is 
modulated by the stellar 
rotation.
The HR 1099 light curve was computed using  \citet{Fekel83} ephemeris:
\begin{equation}
HJD=2442766.080 + 2.83774E
\label{ephemeris}
\end{equation}
All the APT observations included the faint
visual tertiary 
companion HD\,22468B (V$_{\rm ter}$=8.83, \citet{Eggen66}), which is 6 arcsec away from the  
spectroscopic binary. 
Its contribution to the observed light curve {is  $\Delta V=-0.061$ mag. Before performing spot 
modelling, the V-band differential magnitudes were corrected for the third component's contribution:} 
\begin{equation}
\Delta V $=$ 2.5 \log(10^{-0.4 \Delta V_{\mathrm{obs}}} -10^{-0.4(V_{\mathrm{ter}}-V_{\mathrm{comp}})})
\end{equation}
From the light curve maximum V$_{\rm max}=5.744$ at the 1998.89 epoch and 
assuming V $=$ 7.2 
for the  magnitude  of the G5 component \citep{Henry95}, the maximum of the primary K1\,IV was 
estimated to be V $=$ 6.074.}

\section{Results}
\subsection{Radial Velocity and Orbital Solution}
{According to \citet{Fekel83}}, the orbital period has remained {constant} over 60 years. 
Therefore, to compute the observed phases for our analysis, we adopted his 
ephemeris (see Eq.~(1)). The uncertainty in the period (0.00001 d) is small 
enough to ensure that, in the three week  interval of our campaign, any 
possible phase drift is less than 0.001 cycles. We therefore solved for the 
following orbital parameters: $\phi_{0}$ (phase of conjunction with the K1 IV 
sub-giant in front), $\gamma$ (systemic velocity), $K_{pri}, K_{sec}$ 
(semi-amplitudes of the velocity curves for the K1 IV and G5 V star, respectively).
The radial velocities were determined by using a two-Gaussian fitting 
technique to each profile of Fe\,{\sc i} 6430.84 \AA\ and Ca\,{\sc i} 
6439.07 \AA\ observed with HEROS on the {Dutch} 0.9\,m telescope at ESO. This site was the only one that observed 
those photospheric lines during almost all the campaign. The individual 
profiles of each component were resolved and the velocity shifts were measured
with respect to the laboratory wavelengths of the Fe\,{\sc i} and Ca\,{\sc i} lines. 
We obtained altogether {100} new velocities (25 for each component in the two photospheric lines) and used them 
to recompute the orbital parameters of HR 1099. These velocities are 
listed in Table~4. In order to obtain the orbital parameters, the radial velocity 
curves were fitted by a  double-lined spectroscopic binary (SB2) model. 
Observations near phase conjunction between the components were not
used because of line blending. The orbital parameters are listed in Table~5. 
Fig.~\ref{FigRadialVelocity} shows the observed and computed radial velocity 
curves for the two components of HR 1099, along with their  (O-C) residuals.


\begin{figure}[tpb!]
   \centering
\begin{turn}{270}
    \includegraphics[width=5.75cm]{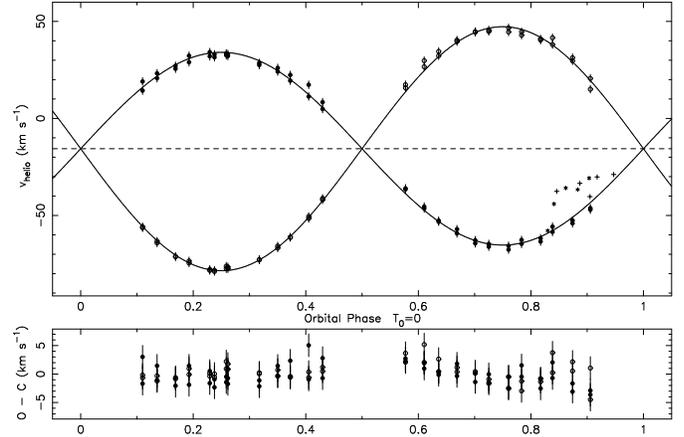}
\end{turn}
   \caption{Observed and computed radial velocity curves for the two components 
   of HR 1099. Measurements for the K1 sub-giant and G5 are {plotted as} 
   filled and open symbols, respectively. Velocities from Fe\,{\sc i} 6430 \AA\ and 
   Ca\,{\sc i} 6439 \AA\ are denoted by coupled symbols. The dashed line is the 
   system's radial velocity. Plus and asterisk symbols show the radial 
   velocities, calculated using H$\alpha$, {during} the optical flares at JD 
   2451145.51 and JD 2451151.07, respectively. The (O-C) velocity residuals are 
   shown in the bottom panel.}
   \label{FigRadialVelocity}%
\end{figure}
\begin{table}[htp!]
\begin{center}
\caption{Orbital elements for the HR 1099 system.}
\begin{tabular}{ll}
\hline
\hline\\
Element (Unit) & Value\\
\hline\\ 
$P_\mathrm{{orb}}$ (days)            &   2.83774 (adopted)\\
T$_{0}$(HJD)           &   2451142.943\, $\pm$\,0.002\\
$\gamma$ (km\,s$^{1}$) & $-15.61$\,$\pm$\,0.20\\
$K_\mathrm{{pri}}$ (km\,s$^{1}$)&   49.66\,$\pm$\,0.34 \\
$K_\mathrm{{sec}}$ (km\,s$^{1}$)&   62.88\,$\pm$\,0.70 \\
$\phi_{0}$             & $-0.0478$ \\
$e$                    &   0.0 (adopted)\\
$a_{1}$\,sin\,$i$ (km)   & 1.94\,$\pm$\,0.01\,x\,10$^{6}$\\
$a_{2}$\,sin\,$i$ (km)   & 2.45\,$\pm$\,0.03\,x\,10$^{6}$\\
$a$\,sin\,$i$ (km)       & 4.39\,$\pm$\,0.03\,x\,10$^{6}$\\
$M_{1}$\,sin$^{3}\,i$\,(M$_{\sun}$) &  0.234\,$\pm$\,0.005 \\
$M_{2}$\,sin$^{3}\,i$\,(M$_{\sun}$) &  0.185\,$\pm$\,0.004 \\
$M_{1}$/$M_{2}$          & 1.266\,$\pm$\,0.011\\
RMS for solution (km\,s$^{-1}$) & 1.85\\
\hline \\
\end{tabular}
\end{center}
\label{TableOrbitalSolution1}
\end{table}
\begin{table}[htb!]
\begin{center}
\caption{A comparison of orbital parameters for HR 1099. $K_{\rm pri}$ and 
$K_{\rm sec}$ denote the velocity amplitudes of the primary and the secondary, 
respectively. $\gamma$ is the systemic velocity, and $\phi_{0}$ is the orbital 
phase of the superior conjunction (primary in front).}
\begin{tabular}{l@{}c@{}c@{}cc}
\hline
\hline\\
Reference & $K_{\mathrm{pri}}$ & $K_\mathrm{{sec}}$ & $\gamma$ & $\phi_{0}$\\
          &           &  (km\,s$^{-1}$)         &          & \\
\hline\\ 
This paper&     49.7    &   62.9     &  $-15.6$        & $-0.0478$\\
\citet{Strass00} & 52.6 & 64.1 & $-15.9$ & $-0.0384$\\
\citet{Donati99}                & 50.0 & 63.1 & $-14.3$ & $-0.0378$\\
\citet{Donati92}         & 50.0 & 62.8 & $-15.4$ & +0.008 \\
\citet{Fekel83}                 & 49.4 & 61.7 & $-15.0$ & 0\\
\hline \\
\end{tabular}
\end{center}
\label{TableOrbitalSolution2}
\end{table}
In Table~6 we compare our derived orbital parameters with previous determinations.
The results of our analysis are in good 
agreement with those obtained by {previous} authors. The value we found 
for $\phi_{0}$ seems to support previous results by \citet{Donati99} and 
\citet{Strass00}, who suggested a slow but  significant variation of  
the orbital phase {at superior  conjunction with time; providing evidence for orbital period variation}. \citet{Applegate92}, and 
most recently \citet{Lanza99}, interpreted such  a variation as a change of the quadrupole-moment 
 of the primary, {along the activity cycle, arising from cyclic exchanges between kinetic and 
magnetic 
energy driven by a hydromagnetic stellar dynamo}. Such a mechanism has been used by
\citet{Kalimeris95} to explain the {orbital period} variations in another RS CVn system: SZ 
Psc. 
\citet{Fekel83} estimated an 
orbital inclination of $33^o\,\pm\,2^o$, this value also used by 
\citet{Vogt83} and \citet{Vogt99}. Although \citet{Donati99} preferred a value 
of $40^o\,\pm\,5^o$, we keep the inclination fixed at $33^o$ for all further 
calculations. Adopting this inclination angle and the  values of  $K_{\rm pri}$ and $K_{\rm sec}$, the 
derived masses of the components of HR 1099 are: 
$M_{1}=1.45M_{\sun}$; $M_{2}=1.14M_{\sun}$.

\subsection{Photometry and Spot Modelling}
{The properties of the photospheric spotted regions or, more precisely, of the total 
spotted area and its distribution versus stellar longitude,  can be obtained by analysing the rotational 
modulation of the optical band flux. However, due to the low information content of wide-band 
photometry, the derived maps are not unique.
In order to obtain unique and stable solutions, some a priori constraints on the properties of the  photospheric maps, that is a  regularization criterion, must be assumed.
In the present study  we used both the Maximum Entropy  method \citep[hereafter ME,][]{Vogt87} 
and the Tikhonov  
criterion \citep[hereafter T,][]{Piskunov90}.

The star's surface is divided into pixels and the specific  intensity of the $i$-th pixel {is assumed 
to be given by
\begin{equation} 
I_i = f_i I_s + (1-f_i) I_u
\end{equation}
where I$_u$ and I$_s$ = C$_s$ I$_u$ are the specific intensity of the unspotted and spotted 
photosphere of the $i$-th pixel (C$_s$ is the brightness contrast which we assume to be 
constant) and  $0 \le f_i \leq 1$ is the pixel fraction covered by spots}. 
The final map is obtained by finding the distribution of $f_i$ that gives a constrained extreme  of the ME 
or T functionals, subject to a $\chi^2$ limit. A detailed description of our modelling approach, 
source of 
errors and accuracy can be found in \citet{Lanza98}.

The numerical code we used to compute the solutions adopts Roche geometry to describe the 
surfaces of 
the components and Kurucz's\footnote{http://cfaku5.harvard.edu} atmospheric models to compute 
the surface fluxes and linear limb-darkening 
in the V-band \citep{Lanza01}. Gravity darkening and reflection effect coefficients were introduced 
according to the procedure outlined by \citet{Lanza98}. 
The stellar and model parameters adopted to compute the final maps are listed in 
Table~\ref{TableStellarParameters}.
\begin{table}
\begin{center}
\caption{Stellar and model parameters of HR 1099 assumed to compute the photospheric maps.}
\begin{tabular}{@{}lccc@{}}
\hline
\hline\\
Stellar parameters                		 & K1\,IV     & G5\,V    & Ref.\\
\hline\\
Brightest V magnitude           		 & 6.074      & 7.2      &  1\\
V-band fractional luminosity     		 & 0.784      & 0.215    &  2 \\
Limb-darkening coeff. (unspotted)           & 0.792      & 0.734    &  3\\
Limb-darkening coeff. (spotted)             & 0.823      & 0.798    &  3\\
C$_{\rm s}$                                      & 0.161      & 0.318    &  3\\
Bolometric correction (mag)      		 & -0.50      & -0.10    &  3\\
Gravity darkening               		 & 0.25       & 0.25     &  4\\
Surface gravity log $g$ (cm sec$^{-2}$)		 & 3.30       & 4.26     &  5\\
Effective temperature (K)        		 & 4750       & 5500     &  6 \\
Starspot temperature (K)        		 & 3750       & 4500     &  7\\
Roche potential                                  & 3.8624     & 19.6412  & 4\\
Fract. radius (point)                            & 0.3572     & 0.1070   & 4\\
Fract. radius (side)                             & 0.3330     & 0.1067   & 4\\
Fract. radius (pole)                             & 0.3225     & 0.1066   & 4\\
Fract. radius (back)                             & 0.3465     & 0.1070   & 4\\

\hline\\

\end{tabular}
\label{TableStellarParameters}
\end{center}
\hspace{.0cm}[1] see Sect.~2.1; [2] present paper; [3] Kurucz's models (2000);\\
\hspace{.0cm}[4] \citet{Kopal89}; [5] according to mass and radius; \\
\hspace{.0cm}[6] \citet{Donati99}; [7] \citet{Strass00}.\\
\end{table}

As shown in Figs.~\ref{FigPhotometry1} \& \ref{FigPhotometry2}, a single large spotted 
region is 
present on the K1\,IV primary, centered around phase 0.85. A  much smaller spotted region 
in the ME map, about 45 degrees apart from the primary, is not resolved in the T solution, which 
represents  the 
smoothest map that fits the light curve.
Spots are also present on the secondary G5 component, centered around phase 0.0 and their area 
is  smaller than on the K1\,IV component.

The resulting spot pattern appears to be concentrated on the northern hemispheres of both components. {This is due to the fact that} 
spots located at latitudes below $\simeq -33^{\circ}$ cannot contribute to the variation of  the 
observed flux because of the low value 
($i\simeq 33^{\circ}$) of the inclination of the rotation axis.
The ME modelling tends to minimize the total spotted area (A$_{\rm ME}=5.49\%$ for K1\,IV, 
A$_{\rm ME}=1.60\%$ for 
G5\,V), giving more compact and darker surface inhomogeneities. The T modelling tends to smooth the 
intensity fluctuations 
correlating the filling factors of nearby pixels and leads to photospheric maps with huge and
smooth spots (A$_{\rm T}=5.65\%$ for K1\,IV, A$_{\rm T}=3.14\%$ for G5\,V). 
\citet{Lanza98} pointed out that the regularized maps should be regarded as an intermediate step of 
the analysis. They must be used to derive quantities not depending on the regularizing criterion 
adopted, that is 
the distribution of the spots vs. longitude and its variation in time. 
Actually, apart from the different total spotted area between the ME and T solutions, both 
regularizing 
criteria reproduce a similar distribution of spotted area versus stellar longitude.

Absolute properties of the spots cannot be extracted from single-band data because systematic 
errors arise 
from the unknown unspotted light level, which we had to assume to coincide with the brightest magnitude at 
the 1998.89 epoch, 
and the assumption of single-temperature spots.

It is interesting to point out that both optical flares detected during the MUSICOS 98 campaign at JD 2451145.51 
and 
JD 2451151.07,  started around phase=0.80, {that is at the rotation phase where  the largest 
active region on the K1\,IV component was facing the observer}. Unfortunately, no simultaneous photometric observations 
were 
obtained at those epochs because of bad weather conditions.}

\subsection{Chromospheric Activity Indicators}
The spectra presented in this paper allow us to study the behaviour of the 
different optical chromospheric activity indicators 
such as: 
Na\,{\sc i} D$_{1}$, D$_{2}$ doublet, (formed in the upper photosphere), 
Ca\,{\sc ii} H \& K lines (lower chromosphere), H$\alpha$, H$\beta$ (middle 
chromosphere), and He\,{\sc i} D$_{3}$ (upper chromosphere), as well as a large 
number of photospheric lines which can in some instances be affected by 
chromospheric activity (i.e. Fe\,{\sc i} 3906.5 \AA, Mn\,{\sc i} 5341.1 \AA, Fe\,{\sc i} 
5430.0 \AA\ .. see \citet{Doyle01}). 

In Figs.~\ref{FigHalphaSerie}-\ref{FigCaiiSerie} we show the time series of the 
Ca\,{\sc ii} H \& K lines, H$\alpha$, H$\beta$, Na\,{\sc i} D$_{1}$, D$_{2}$ 
doublet and He\,{\sc i} D$_{3}$  observed during the MUSICOS 98 
campaign. The vertical axis represents the accumulative orbital phase, {the first orbital revolution beginning at 
zero orbital phase}.

We have measured the equivalent width (EW) for six lines 
with respect to a local continuum 
interpolated by a linear fit to line-free nearby continuum portions included 
within the covered spectral interval.  The
wavelength intervals are 6557.5--6566.5 \AA\ for H$\alpha$, 5883.0--5903.0 \AA\ 
for Na\,{\sc i} D$_{1}$, D$_{2}$ doublet, 5872.0--5877.5 \AA\ for He\,{\sc i} 
D$_{3}$, 4857.5--4865.5 \AA\ for H$\beta$, 3964.5--3972.5 \AA\ for Ca\,{\sc ii} H and
3929.7--3937.7 \AA\ for Ca\,{\sc ii} K.  In this analysis we consider positive
and negative values {of the EW as referring to lines in absorption and emission, respectively.} In 
Fig.~\ref{FigEW} we plot  the 
equivalent width measured for the H$\alpha$, Na\,{\sc i} D$_{1}$, D$_{2}$ 
doublet, He\,{\sc i} D$_{3}$, H$\beta$ , Ca\,{\sc ii} H and
Ca\,{\sc ii} K lines, versus Julian date and the orbital phase.

\begin{figure}[htbp!]
   \centering
   \resizebox{\hsize}{!}{\includegraphics{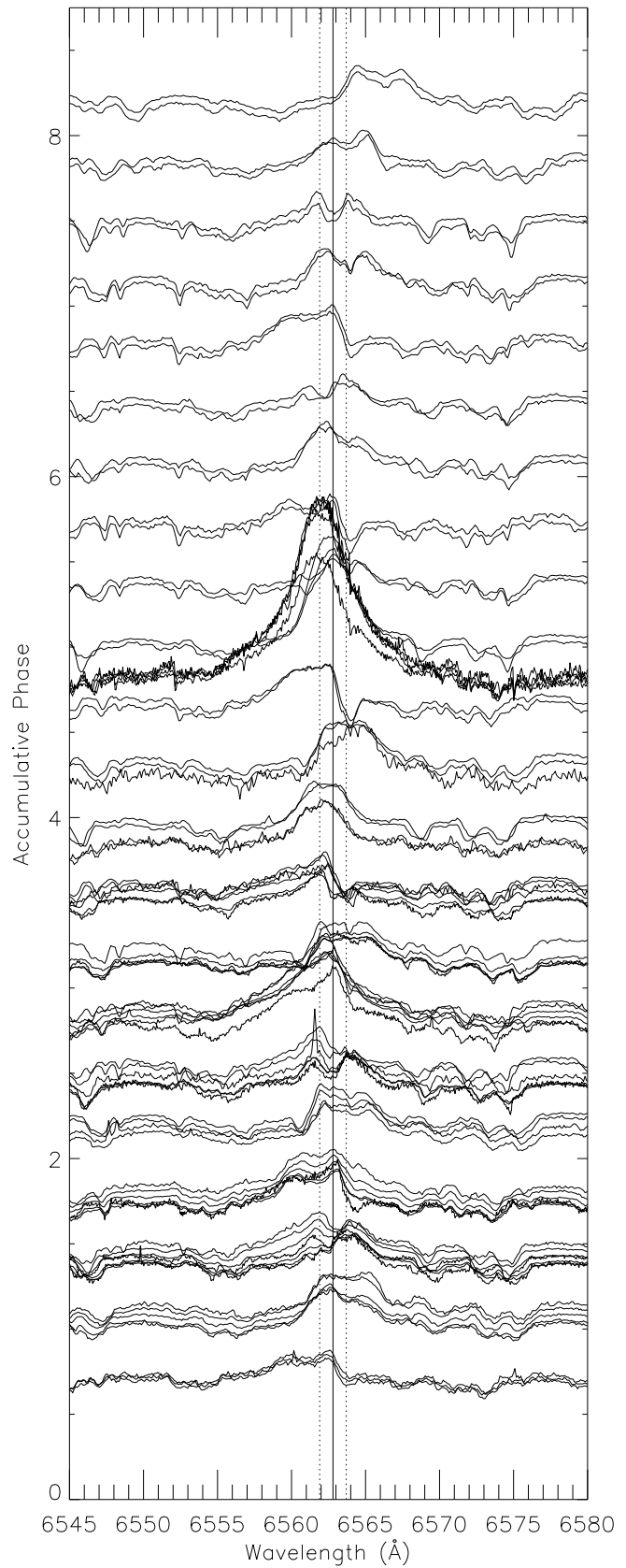}}
\vspace*{0.1cm}
   \caption{Time series of the H$\alpha$ line, obtained during the entire 3-week 
   MUSICOS 98 campaign. {The vertical scale indicates the accumulative orbital phase.} The vertical lines correspond to the rest wavelength
   of the line (solid line) and the corresponding maximum rotational broadening 
   {of the K1\,IV component of}  $v\,\sin\,i =\pm\,41\ \mathrm{km\,s^{-1}}$ (dotted line), 
respectively.}
   \label{FigHalphaSerie}
\end{figure}


\begin{figure}[htbp!]
   \centering
   \resizebox{\hsize}{!}{\includegraphics{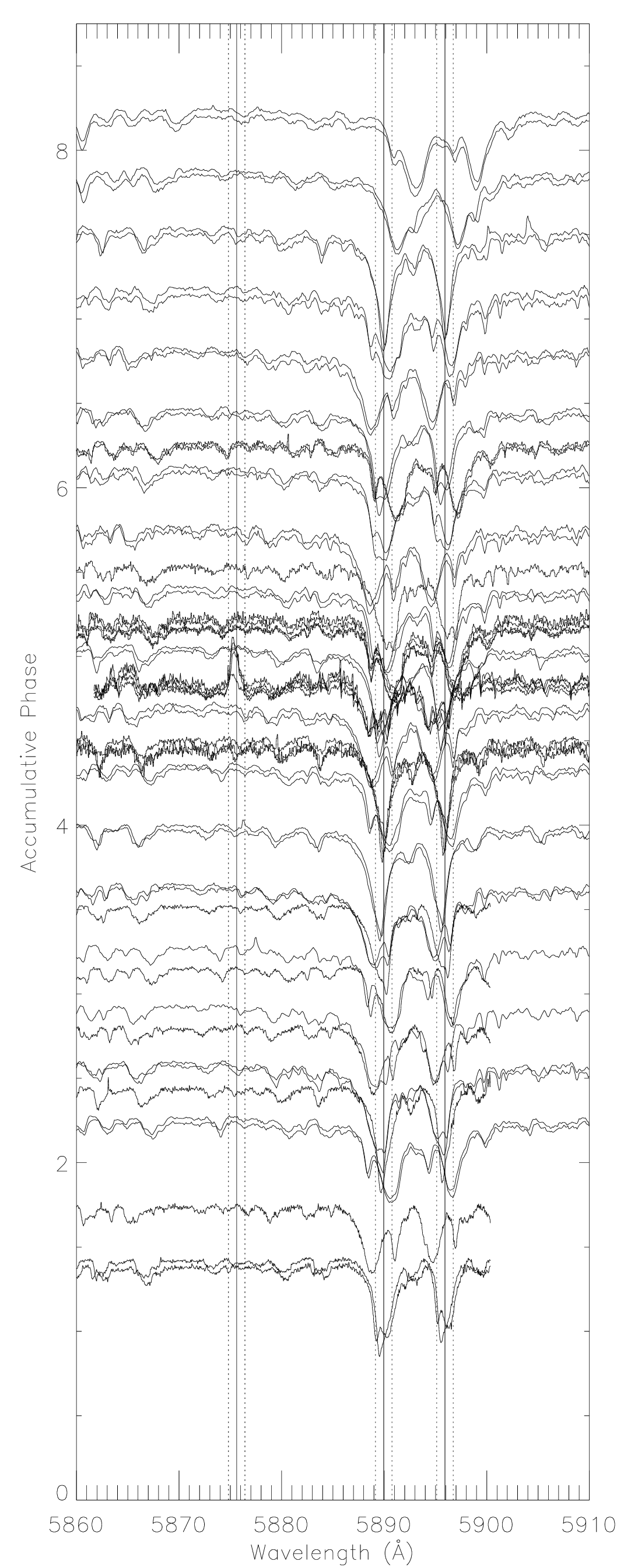}}
\vspace*{0.1cm}
   \caption{Same as Fig.~\ref{FigHalphaSerie}, but for 
   the He\,{\sc i} D$_{3}$  and Na\,{\sc i} D$_{1}$, D$_{2}$ doublet.
   }
   \label{FigHeNaSerie}
\end{figure}


\begin{figure}[htbp!]
   \centering
   \resizebox{\hsize}{!}{\includegraphics{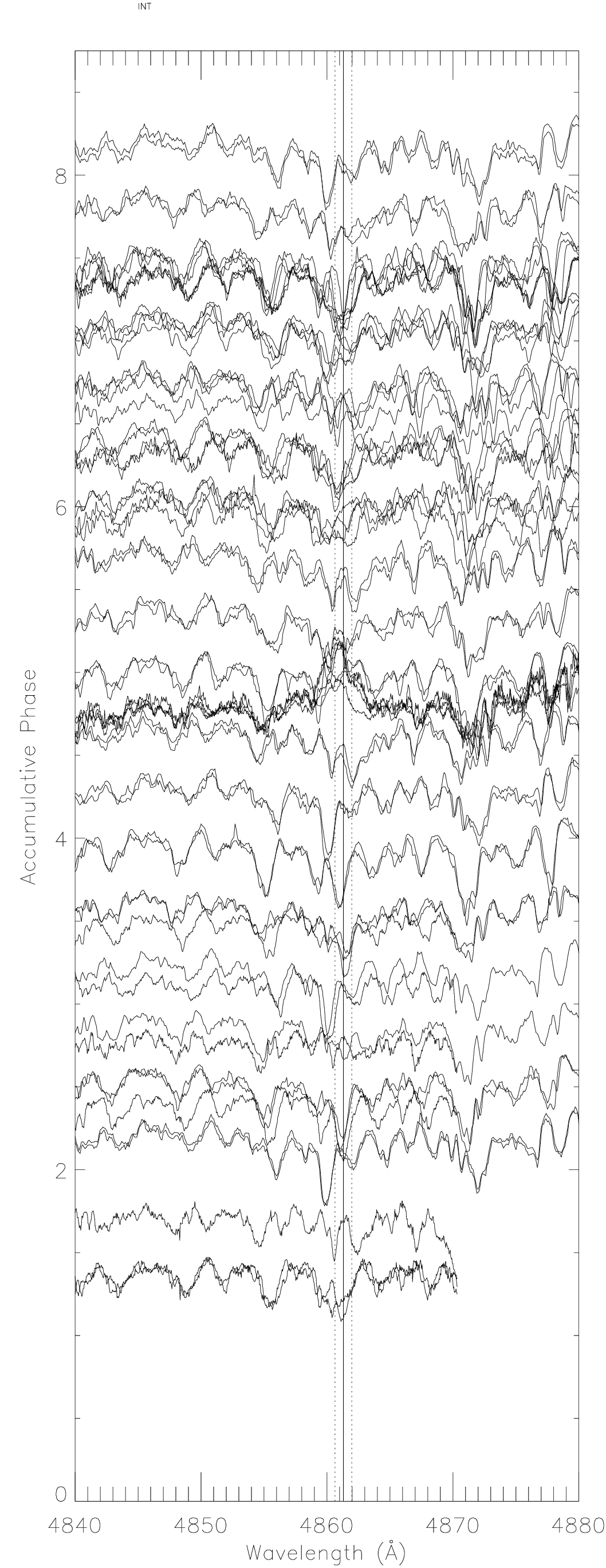}}
\vspace*{0.1cm}
   \caption{Same as Fig.~\ref{FigHalphaSerie}, but for 
   the H$\beta$ 4861 \AA\ line. 
   }
   \label{FigHbetaSerie}
\end{figure}


\begin{figure}[htbp!]
   \centering
   \resizebox{\hsize}{!}{\includegraphics{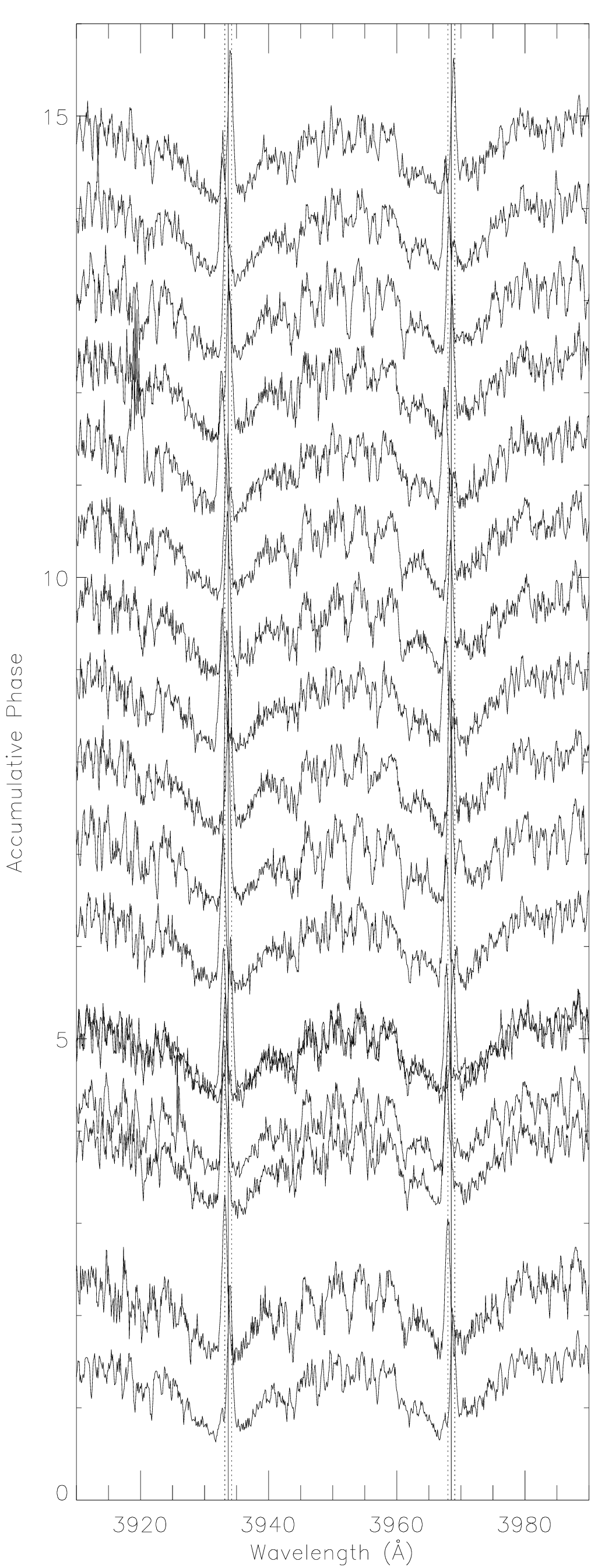}}
\vspace*{0.1cm}
   \caption{Same as Fig.~\ref{FigHalphaSerie}, but for 
   the Ca\,{\sc ii} H \& K lines. 
   Note that the gap between spectra has been doubled for clarity sake.}
   \label{FigCaiiSerie}
\end{figure}

\subsection{Rotational Modulation}
We have observed an increase in emission in the EW of the He\,{\sc i} D$_{3}$ line and the Na\,{\sc i} D$_{1}$, D$_{2}$ 
doublet, between 
$\phi=$0.5 and $\phi=$1.0 (see the second and third right panel of Fig.~\ref{FigEW}). 
{These} changes in EW are consistent with rotational 
modulations in the He\,{\sc i} D$_{3}$ line, that could be attributed to
pumping of the chromospheric emission by coronal X-rays from {an overlying} active 
region. A similar behaviour can be seen in the EW of the H$\alpha$ line with an increase in emission
between $\phi=$0.7 and  $\phi=$1.0 (see the top right panel of Fig.~\ref{FigEW}). 
These enhancements are unlikely to be a temporary, flare-like phenomenon, since they 
{have} been observed for almost nine rotations of the star. {Note that the flares observed during 
the campaign correspond to significantly larger increases in emission in the EW (see Fig.~\ref{FigEW}}). The H$\beta$ and the Ca\,{\sc ii} H \& K lines do not show any obvious 
rotational modulation. Such rotational modulation 
behaviour in chromospheric lines has been observed previously in other RS CVn 
systems \citep{Rodono87,Busa99,Padmakar00}. {It may indicate that the 
distribution of 
chromospheric active regions is ascribed to localised regions and not uniformly distributed over the
stellar surface, which would not produce any rotational signature. This vertical structure would resembles that of solar active 
regions.}

The observed behaviour in the Na\,{\sc i} D$_{1}$, D$_{2}$ 
doublet, the He\,{\sc i} D$_{3}$ and H$\alpha$ 
lines, seems to show a sort of anti-correlation with the optical  light curve
(cf. Figs.~\ref{FigPhotometry1} \& \ref{FigPhotometry2}). It means that the 
chromospheric emission maximum  is in close coincidence with the minimum of the light curve (when the largest spotted region
is facing the observer). 

\section{Flares Analysis}
During the campaign, two optical flares were observed, one at JD 2451145.51 (28-11-98) lasting 
about 0.63 days and a second flare at JD
2451151.07 (03-12-98) lasting about 1.1 days (see Fig.~\ref{FigEW}). 

\subsection{The Hydrogen Balmer lines}

\subsubsection{The H$\alpha$ and H$\beta$ lines}
The emission or filling-in of the H$\alpha$ (6562.8 \AA) and H$\beta$ 
(4861.3 \AA) lines is one of the primary optical indicators 
of chromospheric activity in RS CVn systems. 

The first optical flare, at JD 2451145.51, shows an increase in emission in the measured 
EW by almost a factor of two in H$\alpha$. Broad components are observed at the 
beginning and at the maximum of this  flare (see left panel of 
Fig.~\ref{FigFlare1}). The H$\alpha$ profile at the start of the flare has a full-width half maximum (FWHM) 
of 2.87 \AA\ (equivalent to 131 $\mathrm{km\,s^{-1}}$) and shows 
a noticeable  blue wing excess in four flare 
spectra. The largest FWHM, with a value of 4.52 \AA\ (207 $\mathrm{km\,s^{-1}}$), 
was obtained for the spectrum at $\phi$=0.7981. The spectrum at $\phi$=0.8571, 
which corresponds to the  maximum H$\alpha$ emission, has a FWHM of 3.47 \AA\ (159 
$\mathrm{km\,s^{-1}}$). A narrow absorption feature was also observed in this 
spectrum (see Fig.~\ref{FigFlare1}). Observations of the H$\beta$ line were not 
made during most of the flare, however, the spectrum observed around flare 
maximum, at $\phi$=0.8571, did show a filling-in of the H$\beta$ line (see 
Fig.~\ref{FigEW} and top right panel in Fig.~\ref{FigFlare1}).  

The second optical flare, at JD 2451151.07, shows an increase in emission in the measured 
EW by a factor of four in H$\alpha$. The left panel in Fig.~\ref{FigFlare2} shows 
very broad components at the beginning and at the maximum of this event.
The H$\alpha$ profile at the start of the second flare, $\phi$=0.6203, has a 
FWHM  of 3.13 \AA\ (equivalent to 141 $\mathrm{km\,s^{-1}}$). The largest 
FWHM, with a value of 3.82 \AA\ (175 $\mathrm{km\,s^{-1}}$), was obtained for 
the spectrum at $\phi$=0.8141. The spectrum at $\phi$=0.8354, which corresponds 
to the maximum H$\alpha$ emission, has a FWHM of 3.71 \AA\ (170 
$\mathrm{km\,s^{-1}}$). The spectra before and after flare maximum show
a symmetrical broadening, {with a base width for the blue and red wings,} of 
6.25 \AA\ (286 $\mathrm{km\,s^{-1}}$).
Similar values for the FWHM were found by \citet{Foing94} for another flare in the 
HR 1099 system. During this second optical flare, the H$\beta$ line turns into 
emission (see right panel of Fig.~\ref{FigFlare2}). We measured a FWHM of 
2.33 \AA\ (144 $\mathrm{km\,s^{-1}}$) for the spectrum at $\phi$=0.8144. It is 
slightly larger than the FWHM obtained during the flare maximum at $\phi$=0.8354, 
which was 2.14 \AA\ (132 $\mathrm{km\,s^{-1}}$). The absorption feature in the 
H$\beta$ emission spectra around flare maximum may be due to the 
secondary star.


\begin{figure*}[htbp!]
   \centering
\includegraphics[width=17cm]{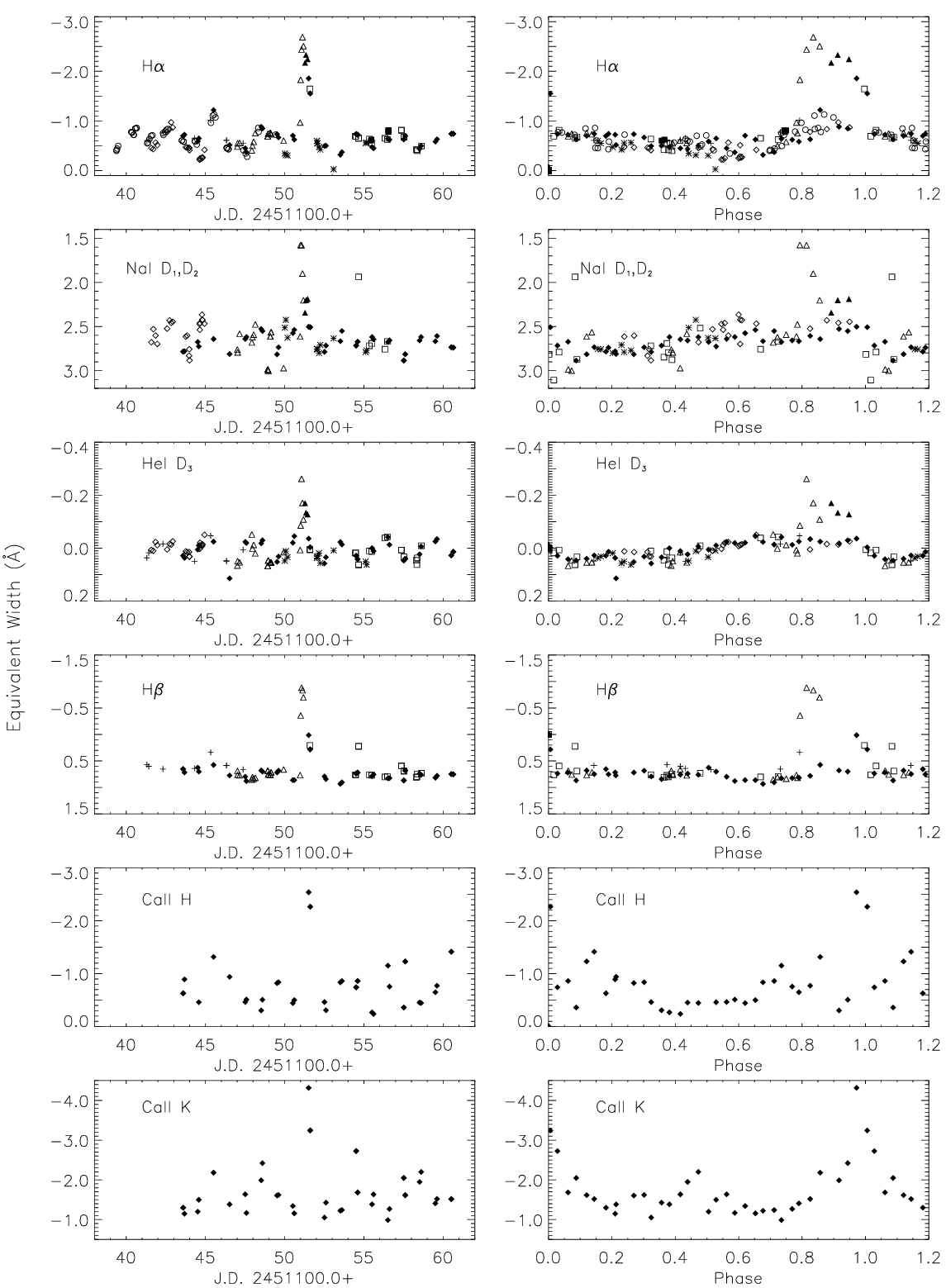}
   \caption{ The equivalent width as a function of Julian date and as a 
   function of phase for H$\alpha$, Na\,{\sc i} D$_{1}$,D$_{2}$ doublet, 
   He\,{\sc i} D$_{3}$, H$\beta$ and Ca\,{\sc ii} H \& K lines observed at the 
different sites, namely: OHP152 (open circle), OHP193 (+), KPNO (open diamond), 
ESO (filled diamond), Mt.Stromlo (open triangle), Xinglong (*), INT (open 
square), LNA (filled square) and SAAO (filled triangle).}
   \label{FigEW}
\end{figure*}

Apart from the two optical flares, H$\alpha$ showed flux enhancements  
around JD 2451143.0, 2451156.5 and 2451157.3. These JDs 
correspond to phases between 0.8 and 0.9. Also two enhancements were 
observed in the H$\beta$ line at around JD 2451155.0 and 2451157.3. {Hereafter, 
we will refer as flare-like to those events which show an enhancement in the average 
line emission although, they may not be proper flares.}

\subsubsection{The Balmer Decrement}
Balmer decrements (flux ratio of higher series members to H$\gamma$) have been 
used to derive plasma densities and temperatures in flare star chromospheres 
\citep{Kunkel70,Gershberg74,Katsova90,Garcia-Alvarez02}. Lacking H$\gamma$ data 
in our spectra, we have instead calculated the Balmer decrement from the 
$\rm{EW_{\mathrm{H\alpha}}/EW_{\mathrm{H\beta}}}$ ratio. During quiescent phases we obtained values around 
1 for the 
$\rm{EW_{\mathrm{H\alpha}}/EW_{\mathrm{H\beta}}}$ ratio, while during the first flare we obtained 
slightly larger values (2-3). During the second 
flare the ratio even reached values of 8. This behaviour shows (unsurprisingly) 
a significant change in the properties of hydrogen emitting regions during the 
flare. {\citet{Kunkel70} showed that this behaviour is possible if the
Balmer lines are driven toward local thermodynamic equilibrium (LTE) conditions
within an emitting region with an electron density of $n_{\rm e}\sim$10$^{13}$
cm$^{-3}$, an electron temperature of $T_{\rm e}\sim$2$\times10^{4}$ K, and a
micro-turbulence velocity $\xi\sim$20 $\mathrm{km\,s^{-1}}$.}


\begin{figure*}[htbp!]
   \centering
  
\includegraphics[width=15cm]{{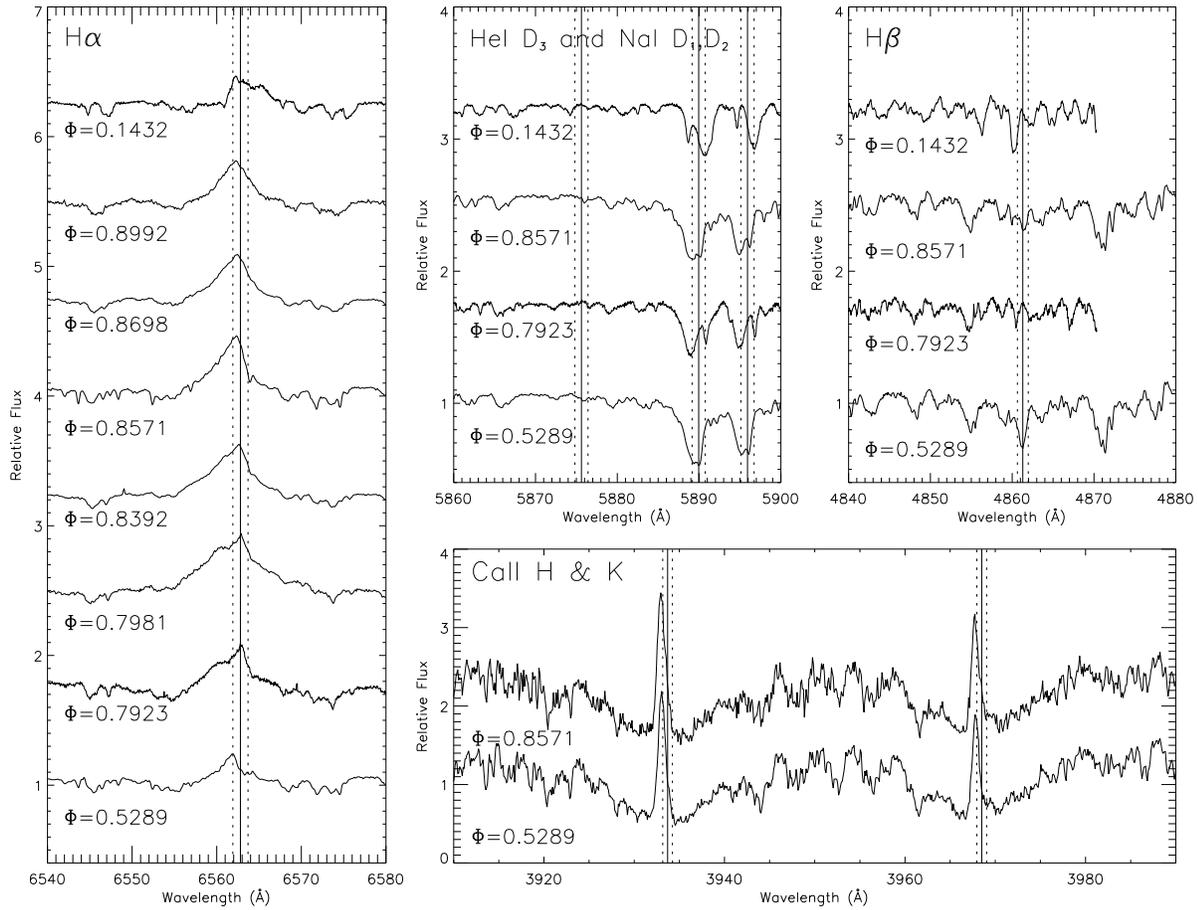}}
   \caption{The observed spectra for
H$\alpha$ (left panel), He\,{\sc i} D$_{3}$, and Na\,{\sc i} D$_{1}$,D$_{2}$ 
doublet (middle panel),  H$\beta$ (upper right panel) {and Ca\,{\sc ii} H \& K (lower right panel)}  of 
the first monitored flare 
starting at JD 2451145.51, arranged in order of the orbital phase. The vertical lines correspond to the rest wavelength
   of the lines (solid line) and the corresponding maximum rotational broadening 
   {of the K1\,IV component of}  $v\,\sin\,i =\pm\,41\ \mathrm{km\,s^{-1}}$ (dotted line), 
respectively.}
   \label{FigFlare1}
\end{figure*}

\subsection{The Na\,{\sc i} D$_{1}$, D$_{2}$ doublet}
The Na\,{\sc i} D$_{1}$ 5895.92 \AA\ and D$_{2}$ 5889.95 \AA\ lines are well 
known temperature and luminosity indicators. These resonance lines are 
collisionally-controlled in the atmospheres of late-type stars, providing  
information about chromospheric activity \citep{Montes97} and for M dwarfs they
have been used to construct model atmospheres \citep{Andretta97}. Unfortunately,
the observations of the 
Na\,{\sc i} D$_{1}$, D$_{2}$ doublet were not made during most of the first 
optical flare. However, the spectra observed around flare maximum, at 
$\phi$=0.8571, did show a slight filling-in (see Fig.~\ref{FigEW} and the top 
middle panel of Fig.~\ref{FigFlare1}). During the second optical flare, the 
Na\,{\sc i} D$_{1}$, D$_{2}$ doublet showed a strong filling-in, which 
during the maximum of this event, turned into emission reversal (see the middle 
panel of Fig.~\ref{FigFlare2}). Another flare-like event seems to have happened at 
around JD 2451155.0.

\subsection{The He\,{\sc i} D$_{3}$ line}
Perhaps the most significant observation in support of flare-like events is the 
detection of emission in the He\,{\sc i} D$_{3}$ line at 5876 \AA. This line has 
a very high excitation level. It has been previously detected in absorption  in active 
stars  and attributed to plages or coronal radiation 
\citep{Huenemoerder86}. It has also been seen in emission during stellar flares 
\citep{Montes97,Montes99,Oliveira99}. In the Sun, it is in emission in the strongest 
flares \citep{Tandberg67}. As for Na\,{\sc i}, observations of He\,{\sc i} 
D$_{3}$ were not 
made during most of the first optical flare. However, the only spectrum observed 
around flare maximum, at $\phi$=0.8571, did not show any apparent change 
(see the top middle panel in Fig.~\ref{FigFlare1}). During 
the second optical flare, the He\,{\sc i} D$_{3}$ line turns into emission (see 
the middle panel of Fig.~\ref{FigFlare2}). We measured a maximum FWHM of 
0.88 \AA\ (45 $\mathrm{km\,s^{-1}}$) for the spectra at $\phi$=0.7983. During 
flare maximum, $\phi$=0.8354, we measured a FWHM of 0.83 \AA\ (42 
$\mathrm{km\,s^{-1}}$). Fig.~\ref{FigEW} shows that, during the second optical 
flare, the He\,{\sc i} D$_{3}$ line peaks before the Balmer lines, in agreement 
with the emission line  behaviour observed in late-type star flares
\citep{Garcia-Alvarez00,Oliveira99}. Three flare-like events seem to happen at around 
JD 2451148.0, 2451156.5 and 2451159.6. These JDs correspond to phases between 
0.8 and 0.9.

\subsection{The Ca\,{\sc ii} H \& K lines}
The Ca\,{\sc ii} H \& K resonance lines have long been the traditional 
diagnostic of chromospheric activity in cool stars since they were
studied by \citet{Wilson78}. Observations of the Ca\,{\sc ii} H \& K lines were 
not made during most of the first optical flare. However, the spectra observed 
around the maximum during both flares, did show a small increase in 
flux (see the bottom panels of Fig.~\ref{FigEW} and Fig.~\ref{FigFlare1}). 
Note that the Ca\,{\sc ii} H \& K lines peaked later 
than the Balmer lines, at around $\phi$=0.95. This behaviour  has been observed 
in flares on late-type stars  \citep{Rodono88, Garcia-Alvarez00}.


\begin{figure*}[htbp!]
   \centering
   \includegraphics[width=15cm]{{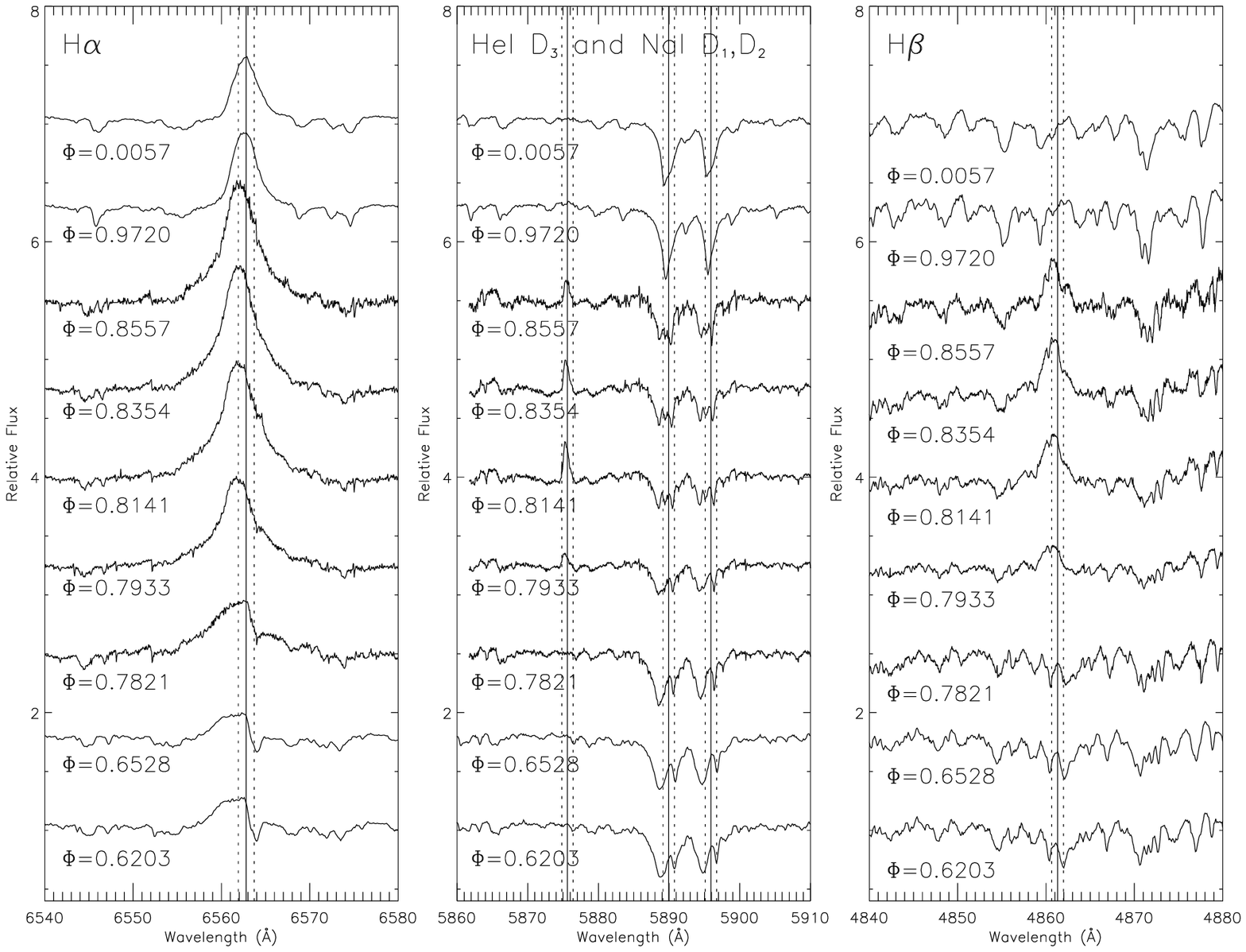}}
   \caption{The observed spectra for
H$\alpha$ (left panel), He\,{\sc i} D$_{3}$ and Na\,{\sc i} D$_{1}$,D$_{2}$ 
doublet (middle panel) and H$\beta$ (right panel) of the second monitored flare
starting at JD 2451151.07 arranged in order of the orbital phase. The vertical lines correspond to the rest wavelength
   of the lines (solid line) and the corresponding maximum rotational broadening 
   {of the K1\,IV component of}  $v\,\sin\,i =\pm\,41\ \mathrm{km\,s^{-1}}$ (dotted line), 
respectively.}
   \label{FigFlare2}
\end{figure*}

\subsection{Flare Location}

In Fig.~\ref{FigRadialVelocity}, together with the radial velocity calculated 
using the photospheric lines, we have plotted the radial velocity for the 
two monitored flares. These radial velocities have been calculated by fitting a 
single Gaussian to the H$\alpha$ profile. The velocity shifts were measured
with respect to the rest wavelength of the H$\alpha$ line. Despite the broad 
component developed by the H$\alpha$ line during flares, this seems to be a 
reliable fit. As a result, we notice that during both flares, the radial 
velocity is slightly displaced (15-20 $\mathrm{km\ s^{-1}}$) compared to the 
center of gravity of the primary. 
This could arise if both flares {took place off the disk of the primary star}.

Note that both flares took place at around the same phase (0.85), but $\sim$6 
days apart suggesting a link between them and the active region {complex, which was crossing the 
central meridian around that phase.}
{ The largest photospheric active region on the K1\,IV component was indeed centered 
around $\phi$=0.85  
(see Fig.~\ref{FigPhotometry1} and Fig.~\ref{FigPhotometry2}). {This behaviour is consistent  with the proposed
link between  
flares and surface active regions}. During the MUSICOS 89 campaign, \citet{Foing94} reported on a 
flare that
took place again at a similar phase (0.87), suggesting that, if this is indeed the
same active region, it is a long-lived feature. \citet{Vogt99} analysed Doppler
Images of HR 1099 and also proposed similar long-lived active regions.


\begin{table}[htbp!]
\begin{center}
\caption{The energy released per exposure (in units of $10^{33}$ ergs) by the 
chromospheric lines during the optical flares observed on HR 1099 in the 
MUSICOS 98 campaign.}
\scriptsize{
\begin{tabular}{cccccc}
\hline
\hline\\
JD & phase & H$\alpha$  & Na\,{\sc i} D$_{1}$,D$_{2}$  & He\,{\sc i} D$_{3}$  & 
H$\beta$  \\
2451100.0+ &   & & & &  \\
\hline\\ 
FLARE 1 & 	 &        &        &	   &        \\
45.328  & 0.7923 &  0.92  &  --    & 0.22  & 2.19   \\
45.345  & 0.7981 &  0.50  &  --    & --	   & --	    \\
45.462  & 0.8392 &  1.42  &  --    & --	   & --	    \\
45.513  & 0.8571 &  2.90  &  0.06  & 0.17  & 1.60   \\
45.549  & 0.8698 &  1.52  &  --    &	-- & --	     \\
45.632  & 0.8992 &  1.25  & --     & --	   & --	     \\\\  
FLARE 2 &        &	  &        &       &	     \\
50.975  & 0.7821 &  0.62  &  0.14  & 0.03  &   1.21      \\
51.007  & 0.7933 &  3.32  &  3.50  & 0.28  &   3.85 	     \\
51.066  & 0.8141 &  5.29  &  3.49  & 0.85  &   5.59 	     \\
51.126  & 0.8354 &  6.09  &  2.44  & 0.56  &   5.43 	     \\
51.184  & 0.8557 &  5.51  &  1.46  & 0.35  &   4.98 	     \\
\hline \\
\end{tabular}
}
\end{center}
\end{table}

\subsection{Energy released}
In order to estimate the energy released in the observed optical chromospheric 
lines during both optical flares, we converted the excess EW (flare EW minus quiet EW of the studied lines) into absolute surface 
fluxes and luminosities. Since we have not observed the entire flare, and several important lines are
missing  (e.g., Ly$\alpha$, Mg\,{\sc ii} h\&k, He\,{\sc 
i} 10830 \AA), our estimates are only lower limits to the total flare energy in 
the chromospheric lines. We have used the calibration of Hall (1996) to obtain 
the stellar continuum flux in the H$\alpha$ region as a function of ($B - V$) and 
then converted the EW into absolute surface flux. For the other lines, we have 
used the continuum flux at H$\alpha$ corrected for  the 
continuum flux ratio $F_{\lambda6563}/F_{\lambda}$. This $F_{\lambda}$ is given by a 
blackbody, with contribution of both the cool and the hot component of HR 1099 at
$T_\mathrm{eff}=4\,840\ \mathrm{K}$ and $T_\mathrm{eff}=5\,460\ \mathrm{K}$, 
respectively. The contribution factors for each component has been obtained 
following the method by \citet{Montes95}. We converted these fluxes into 
luminosities using the radius $R=3.9\,R_{\sun}$, since we 
have assumed that both optical flares took place on the K1\,IV component. 
Table~8 shows the energy released by the chromospheric optical lines for both 
flares. We obtained the flare energies by integrating the measured fluxes in time.
{Adding the energies for all the observed chromospheric lines during each flare, we obtained 
a total flare energy of $1.3 \ 10^{34}$ erg for the first flare and
$5.5 \ 10^{34}$ erg for the second flare, integrating over a period of $\sim$7.3 and 
$\sim$5.5 hours respectively. These values for the energy released are comparable to 
other RS CVn flares \citep{Doyle91,Foing94}. } 


\section{X-ray data}
X-ray observations of HR 1099, with ASM on board RXTE, were carried out at the 
same time as the MUSICOS 98 campaign took place. In Fig.~\ref{FXray} we show 
the X-ray light curve of HR 1099 obtained by RXTE in the S band. 


\begin{figure*}[htbp]
   \centering
   \includegraphics[width=18cm]{{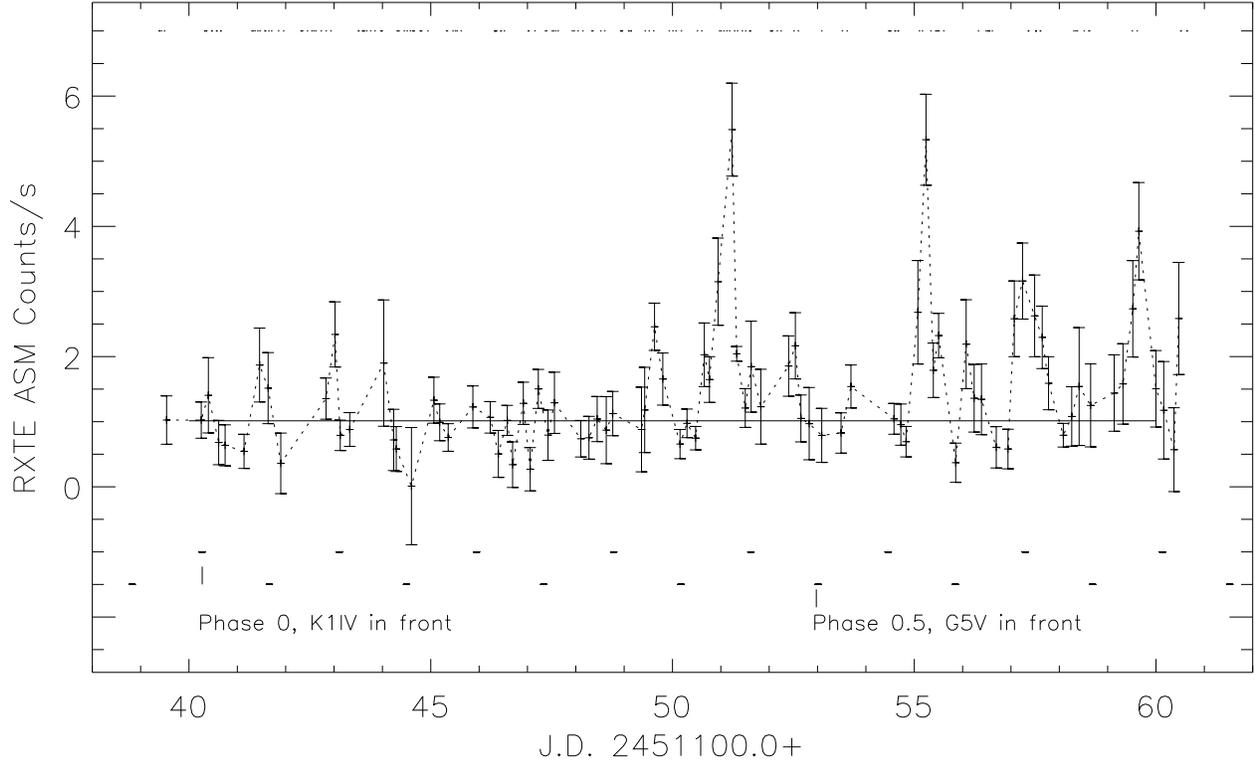}}
   \caption{X-ray light curve observations of HR 1099, from the ASM instrument on 
RXTE satellite, obtained at the same time as the MUSICOS 98 campaign. The band S 
(1.5-12 keV) as a function of Julian date is shown. The spectroscopic optical observations during 
MUSICOS 98 campaign are shown on top of the X-ray data.}
   \label{FXray}
\end{figure*}

\subsection{Long- and Short-Term Variability}
The long-term light curve displays evident variability on short and long 
time-scales. We can divide the RXTE observations into three intervals, according
to the observed activity levels. The 
X-ray data show a medium variability {level} between JD 2451139 
and JD 2451145. In this first interval, which covered around two rotations, we see
three possible flare-like events. A interval of low variability follows, 
covering approximately 1.5 rotations, between  JD 2451145 
and JD 2451149. Finally, the third interval, between JD 2451149 
and JD 2451160, covering four rotations, exhibits strong variability. During this third term six 
possible 
X-ray flare-like events were observed. The flare-like 
events observed in the first and third intervals were only detected in 
the S and A bands but not in the B and the C bands. This suggests that these X-ray events were
soft. 

\subsection{X-ray and Optical Flare Correlation}
We have found that most of the optical flares show a X-ray counterpart. 
During the X-ray event at JD 2451143.02, a small increase in the 
emission of the H$\alpha$ was seen (see the top left panel of 
Fig.~\ref{FigEW}), as well as a filling-in in the Na\,{\sc i} D$_{1}$, D$_{2}$ doublet. 
The optical flare observed at JD 2451151.07 
was also observed in the X-ray range. The X-ray flare observed at JD 2451155.24 
shows an optical counterpart in H$\beta$ line (see Fig.~\ref{FigEW}) and filling-in of the Na\,{\sc i} D$_{1}$, D$_{2}$ 
doublet, although no clear variation was seen in either H$\alpha$ or He\,{\sc i} D$_{3}$.

The flare-like event at 
JD 2451156.74 shows a small increase in the emission of H$\alpha$ and a small filling-in of the 
He\,{\sc i} D$_{3}$ line (see Fig.~\ref{FigEW}). This enhancement was also observed in 
Ca\,{\sc ii} H (not in K) suggesting that it
reflects variability from H$\epsilon$. The possible flare observed at JD 2451157.24 shows  
a small increase in the H$\alpha$ and Ca\,{\sc ii} H \& K lines with also a small 
filling-in of the H$\beta$ line, but no changes
are observed in the other lines. Conversely, the optical flare at JD 
2451145.51 was not observed in any of the X-ray bands.

\subsection{Flare Periodicity}
We noted that most of the X-ray flares and flare-like events observed, took place either 
at $\phi\sim$0.31 (2451149.63 d, 2451152.54 
d, and 2451155.24 d) or at $\phi\sim$0.91 (2451143.02 d, 2451151.23 d, 
2451157.24 d and 2451159.64 d). Based on this fact one could say that there were 
two active regions, which flared during the same epoch. The flares of each active region  show a rotation periodicity which lasted for almost three 
consecutive orbits, from  JD 2451149 to JD 2451158, during what we have called the 
strong variability interval. 

{The possibility of rather persistent active longitudes in the photosphere is well known in the solar case. 
\citet{leto97} found evidence of a frequently flaring and persistent active longitude on EV Lac. 
This, coupled with rotation, may give rise to spurious periodicities or abnormally long-lasting flare events.}
Note that the X-ray flare-like event which would 
have been expected at JD 2451154.2, coupling with the one at JD 2451152.54, is not covered by X-ray observations. 
However, increases in the H$\beta$ and 
Ca\,{\sc ii} H \& K lines were observed {at that epoch}.

{The observed rotational periodicity could have been  caused by two very large flares, 
one on each active region, lasting for three consecutive orbits, which means that  we 
have observed flares which lasted more than eight days!. This kind of long duration 
flares on active RS CVn stars has been previously observed by 
\citet{Montes98}. However, one would expect an a exponential decay in the X-ray flux, 
which was not observed, and a much larger X-ray flux maximum, which was not observed 
either. So we discard a possible very long duration flare as an explanation for the 
observe X-ray light curve. Another possibility is suggested by the behaviour of the 
light curve at JD 
2451150, JD 2451153 and JD 2451156. The X-ray flux appears to peak at 
$\phi\approx$0.31, after which it diminishes until it reaches a minimum  when the G5\,V star is in 
front of the active K1\,IV star ($\phi$=0.5), in 
agreement with other authors \citep{Drake94,Audard01,Sanz-Forcada02}. Similarly, 
\citet{Foing94}, observed a sudden flux decrease in an optical flare, that they suggest was caused 
by the
occultation of the flaring active region, due to the rotation of the system. However, as a 
consequence of X-rays being optically thin, the rotational modulation they might produce 
would modulate the flare in an on-off fashion, but would not produce spiky flares as we observe 
in the X-ray light curve. 
Another possibility, is that we are looking at periodic flaring. 
Several authors have searched for such periodicity with mixed results 
\citep[e.g.][]{Lukatskaya76,Pettersen83, Doyle90b,
Mullan95,Mavridis95}. The confirmation of such a periodicity would support the idea of a 
``flare reservoir'', in other words, an active region with a certain amount of 
stored energy which is then released either as a single flare or as multiple of 
smaller events. \citet{Mullan95} reported X-ray periodicities in late-type flare stars. 
They found that the observations were
consistent with the hypothesis that resonant absorption of MHD waves were occurring at certain 
times in coronal loops. However,
the preferred frequency inferred by \citet{Mullan95} are much lower than the one we observed in 
HR 1099. The larger-scale
magnetic structure obtained in the photospheric maps, around phase 0.85, and the periodicity of the
flares might imply that the reheating events of the same magnetic loop originate in an
interaction between the star and some external trigger, explaining partially the observed X-ray periodicity.}

{We believe that the two active regions, flaring at the same epoch, are
responsible for the observed behaviour of the X-ray light curve. 
Following \citet{Doyle90b}, the probability that the 
three flares arise out of a chance process is rather small ($\leq0.5\%$ and 
$\leq0.6\%$ for the active region at $\phi$=0.31 and at $\phi$=0.91, 
respectively). We conclude that the flares for each assumed
active region show significant periodicity, lasting almost three consecutive orbits 
(from JD 2451149 to JD 2451158). \citet{Doyle90b} 
suggested that the different flares they observed on the eclipsing binary YY Gem originated 
in the same active region close to the stellar surface and were caused by the same trigger 
mechanism. In the case of YY Gem, the explanation was given in terms of filament oscillating 
with a period of 48 mins. It is unlikely that the 2.8 day interval observed here could be 
explained by a similar mechanism, and thus it is more likely that the different groups of 
flares orginate from the
same active region complex. The radial velocities measured for both large optical flares 
discussed in Sect.~4, indicates that they originate from a region  
on the K1 IV component, thus providing some support for this scenario.}

\section{Conclusions}
We have observed the binary system HR 1099 continuously for more than seven 
orbits in the optical spectral range (at high spectral resolution over a wide 
wavelength domain). Contemporary photometric and X-ray observations were performed during 
the campaign.

{Photospheric maps of both components obtained by using Maximum Entropy and Tikhonov 
regularizing criteria show the K1\,IV primary to be the most active with a large spotted 
region in the northern hemisphere, centered around phase 0.85. The detection of optical 
flare events around the same phase is in good agreement with a spatial link between flares and 
active regions.}

We have determined radial velocity curves making use of photospheric lines. The orbital 
parameters were obtained by
fitting the radial velocity 
curves with a double-lined spectroscopic binary (SB2) model. The results of this analysis are in good agreement with
previously published results. The value we found for $\phi_{0}$ seems to {support a slow but 
significant variation in the orbital phase at  the first conjunction with time, {i.e. a change of the orbital period}. It can 
be explained as a consequence of the cyclic variation of the quadrupole-moment 
of the primary  along the magnetic activity cycle. 

Two large optical flares were monitored during the campaign, one of them lasting  more 
than one day. From the flare analysis, we observe that, although both flares
show an increase in H$\alpha$ emission with a broad component, the second 
flare produced a bigger increase in intensity and lasted longer. We have also 
observed that the first flare showed filling-in of H$\beta$, which turned into emission during the 
second 
flare. The He\,{\sc i} D$_{3}$ line was seen in emission
during the second flare. The Na\,{\sc i} 
D$_{1}$, D$_{2}$ doublet showed a filling-in in both flares, being stronger 
during the second one. A similar behaviour was seen in the Ca\,{\sc ii} H \& K lines. We found 
a lower limit to the total flare energy of $1.3 \ 10^{34}$ erg and
$5.5 \ 10^{34}$ erg for the first and second flare, respectively, comparable to other RS CVn 
flares. {We noticed that the He\,{\sc i} D$_{3}$ line was only seen in emission in the 
second, and larger, flare. This is due to the fact that the helium lines are produced under 
generally higher excitation conditions than other chromospheric lines and they are usually not 
detected except in the strongest events. The change of behaviour in the H$\beta$ line, from weak 
absorption in the first flare to strong emission in the second flare was also due that the latter 
was strongest.} 
Note that both flares took place at around the same phase (0.85), but $\sim$6 
days apart suggesting a link between them and the 
large photospheric active region on the K1\,IV component, obtained from the  photospheric spot 
modelling, which is centered 
around the same orbital phase.

{We have detected rotational modulation in the H$\alpha$ and He\,{\sc i} D$_{3}$ lines 
and in the Na\,{\sc i} D$_{1}$, D$_{2}$ doublet, that may indicate non axi-symmetry in the 
distribution of coronal active regions. In other words, this rotational modulation may be due 
to the emission of an active region (or regions), which goes on and off as the star rotates. 
Lines formed at different heights in the atmosphere, would be affected, in a different way, 
by the active region. This will produce rotational modulation for certain lines but not 
necessarily for all of them. However, one has to bear in mind that a group of active regions 
uniformly distributed over the stellar surface would not produce any rotational signature.}

X-ray observations were performed with ASM on board RXTE. 
We have observed clear variability on short and long 
time-scales. A number of flares and flare-like events were detected, some of which correlated 
well with the optical observations. We noted that most of the X-ray events observed, took place 
either at $\phi\sim$0.31 or at $\phi\sim$0.91. 

We plan to compare these results with Doppler Imaging based on the photospheric 
lines, to study the connection between spots, chromospheric emission and flares.

\begin{acknowledgements}
We wish to thanks all those who have contributed to the MUSICOS 98 campaign. 
Research at Armagh Observatory is grant-aided by the Department of Culture, 
Arts and Leisure for Northern Ireland. DGA wishes to thank the Space Science 
Department at ESTEC  for  financial support. DM is supported by the Spanish 
Programa Nacional de Astronom\'{\i}a y Astrof\'{\i}sica, under grant AYA2001-1448. 
JMO research work was supported by the Praxis XXI grant BD9577/96 from the {\it
Funda\c{c}\~{a}o para a Ci\^{e}ncia e a Tecnologia}, Portugal. This 
paper made use of quick look data provided by the RXTE ASM team at MIT and 
GSFC. Stellar activity research at Catania Astrophysical Observatory of the National Institute of 
Astrophysics and at 
Department of Physics and Astronomy of Catania University is supported by the 
MIUR {\it ``Ministero dell'Istruzione, Universit\'a e Ricerca"} 
and by the  {\it Regione Sicilia} that are 
gratefully acknowledged. We also thank the referee M.~Guedel for helpful comments.
\end{acknowledgements}

\bibliographystyle{apj} 
\bibliography{references}

\end{document}